%% file: ms.tex
\documentclass[useAMS,usenatbib]{mn2e}
\usepackage{times,psfig,epsfig,graphics,graphicx,multicol,amssymb,paralist,natbib,threeparttable}
\def\lesssim{\mathrel{\hbox{\rlap{\hbox{\lower4pt\hbox{$\sim$}}}\hbox{$<$}}}}
\def\gesssim{\mathrel{\hbox{\rlap{\hbox{\lower4pt\hbox{$\sim$}}}\hbox{$>$}}}}

\begin{document}  
\title[Mass estimate of MS2137]
  {X-ray and strong lensing mass estimate of MS2137.3--2353}
\author[A. Donnarumma et al.]
  {A.~Donnarumma${}^{1,2,3}$\thanks{E-mail: annamaria.donnarumm2@unibo.it}, S.~Ettori${}^{2,3}$, M.~Meneghetti${}^{2,3}$, L.~Moscardini${}^{1,3}$ \\
$^{1}$ Dipartimento di Astronomia, Universit\`a di Bologna, via Ranzani 1,
I-40127 Bologna, Italy\\
$^{2}$ INAF-Osservatorio Astronomico di Bologna, via Ranzani 1, I-40127 Bologna, Italy\\
$^{3}$ INFN, Sezione di Bologna, viale Berti Pichat 6/2,
I-40127 Bologna, Italy}

\date{Accepted ???. Received ???}

%\pagerange{\pageref{firstpage}--\pageref{lastpage}} \pubyear{2007}

\maketitle

\begin{abstract}
We present new mass estimates of the  cluster of galaxies MS2137.3--2353, inferred from X-ray and strong lensing analyses. This  cluster exhibits  an outstanding strong lensing configuration  and indicates  a  well-relaxed dynamical state, being  most suitable for a  mass reconstruction which combines both techniques. Despite  this, several previous studies have claimed a significant discrepancy between the X-ray and the strong lensing mass estimates. The primary aim of  this paper is to address and explain this mismatch. \\ For this purpose, we have analysed \textit{Chandra} observations to recover the profiles of the intra-cluster medium properties and, assuming a functional form for the matter density, the total mass distribution.  The notable strong--lensing features of MS2137.3 allow us  to reconstruct its  projected  mass in the central regions with  good accuracy, by  taking advantage of the lensing inversion code \emph{Lenstool}. We compare the results obtained for both methods. Our  mass estimates for MS2137.3 are in agreement within  errors, leading to a mean, extrapolated value of $\rm M_{200} \simeq 4.4\pm 0.3 \times  10^{14}\ \rm M_{\odot}$, under the assumption of the Navarro-Frenk-White (NFW)  mass profile. However, the strong lensing mass estimate is affected  by the details of the BCG mass modeling, since the radial arc is a very sensitive probe of  the total mass derivative in the central region.  In particular,  we do not find evidence  for a high concentration for the NFW density profile, as reported in some earlier works. \\
\end{abstract}

\section{Introduction}
According to the hierarchical growth of cosmic structures implied by the concordance model, clusters of galaxies are the most recent gravitationally-relaxed structures.  They are ideal cosmological tools  \citep*[see e.g.][]{voit2005,borgani2006,tozzi2007,diaferio2008}, provided we have a reliable measument of their masses. 
 The most promising techniques for estimating cluster masses are  via X-ray and gravitational lensing analyses. \\ Cluster X-ray luminosity is mainly due to diffuse plasma emission:   the deep potential wells produced by the dark matter component heat  the intra-cluster medium (ICM) to temperatures of some keV, at which point  X-ray photons are emitted through thermal bremsstrahlung. The X-ray emissivity is proportional to the square of  the gas density, thus  it is an excellent tracer of the three--dimensional cluster potential. However, X-ray measurements of  cluster masses imply the assumption of hydrostatic equilibrium of the ICM with the dark matter potential and of spherical symmetry of the cluster mass distribution. Hence the total mass profile can be inferred from the radial profiles of temperature and gas density. \\ On the contrary, the gravitational lensing effect, i.e. the light deflection due to the space--time curvature induced by massive objects, allows for the  determination of the projected surface mass density  of the lens, regardless of its dynamical state or the nature of the intervening  matter. However, this effect is determined by all massive structures along the line of sight, so lensing mass measurements are subject to foreground and background contaminations. Moreover, the lensing effect is very sensitive to features of the mass distribution such as its ellipticity and asymmetry, as well as to the presence of substructures \citep{meneghetti2007b}, which complicate the mass reconstruction. \\The X-ray and lensing methods are clearly complementary, allowing in principle to combine 2- and 3-dimensional constraints to obtain a reliable mass estimate (see for example \citealp{allen1998,ettori2003,bradac2008}); a limit to such a comparative analysis is the significant disagreement between strong lensing and X-ray mass estimates claimed in the literature (\citealp{wu1996,smail1997,ota2004,voigt2006,gitti2007,halkola2008}). Many convincing explanations have been suggested, but to date, the discrepancy between X-ray and strong lensing results  appears to be an issue. \\ 
Moreover, galaxy clusters enable one to determine dark matter (DM) halo density profiles, being mainly composed by dark matter. As a consequence, they could help in  verifying the CDM model predictions for the formation of cosmic structures: for example, if DM density profiles can be approximated on all mass scales by ``universal'' profiles, such as the  Navarro-Frenk-White (NFW) profile \citep*{navarro1995,navarro1997}, as predicted by N--body simulations of structure formation in hierarchical collapse models.
In particular, the inner DM density slope is a very debated issue: several studies  have argued that it could deviate from the NFW predictions \citep{moore1998,jing2000,gao2008}. Galaxy clusters play a crucial role in solving this issue, since they  allow to distinguish between cuspy and flat profiles due to their low concentrations.   An emblematic case in both of these debates is MS2137.3-2353: even though this cluster has a relaxed appearence, there is still disagreement both regarding its inner slope value \citep{dalal2003,sand2004,gavazzi2005a,sand2008} and the X-ray and strong lensing mass estimates \citep{gavazzi2003,gavazzi2005,comerford2006,comerford2007}. \\
In this work we will focus on the latter topic. We present  new X-ray and strong lensing mass estimates, based respectively on a non-parametric analysis of gas density and temperature profiles through \textit{Chandra} observations,  and on a strong lensing reconstruction,  performed using the \emph{Lenstool} analysis software  \citep{kneib1996,jullo2007}.
\\ 
This paper is organised as follows. In $\S\ $\ref{sec:xray} we will discuss our X-ray analysis, focusing on data reduction and on the method applied to recover the total and gas mass profiles, and we will summarize our X-ray results.  The strong lensing analysis is presented in $\S\ $\ref{sec:sl},  where we will briefly discuss our main findings. We will compare the X-ray and strong lensing results in $\S\ $\ref{sec:comp}: a comparison with previous analyses can be found in $\S\ $\ref{sec:prev_comp}. Finally, we will summarize our results and draw our conclusions in $\S\ $\ref{sec:concl}. \\
Throughout this work we assume a flat $\Lambda$CDM cosmology, with matter density parameter $\Omega_{\rm m}$=0.3, cosmological constant density parameter $\Omega_{\Lambda}$=0.7, and Hubble constant $H_{0}=70 \ \rm  h^{-1} \rm  km\ s^{-1} Mpc^{-1}$. Unless otherwise stated, all uncertainties are referred to a $68\%$ confidence level.

\label{sec:intro}

\section{X-ray analysis}
\label{sec:xray}
\subsection{X-ray data reduction}
MS2137.3--2353 is a rich, X-ray luminous cluster at $z \!=\!0.313$ \citep{stocke1991}: within a radius of $\simeq 0.7$ Mpc, its luminosity in the [0.5--8.0] keV band\footnote{The X-ray luminosity was derived fitting the cluster integrated spectra in the radial range $[0 - 0.7]$ Mpc using the \emph{XSPEC} analysis package.}  is  $\simeq 1.4 \times 10^{45}\ \rm erg \ s^{-1}$. 
 This cluster exhibits several indications of a well relaxed dynamical state, for instance the absence of evident substructures and a central X-ray surface brightness peak, associated with a cool core. These characteristics support the hypothesis of hydrostatic equilibrium underlying our X-ray mass estimate. 

 \begin{figure}
\psfig{figure=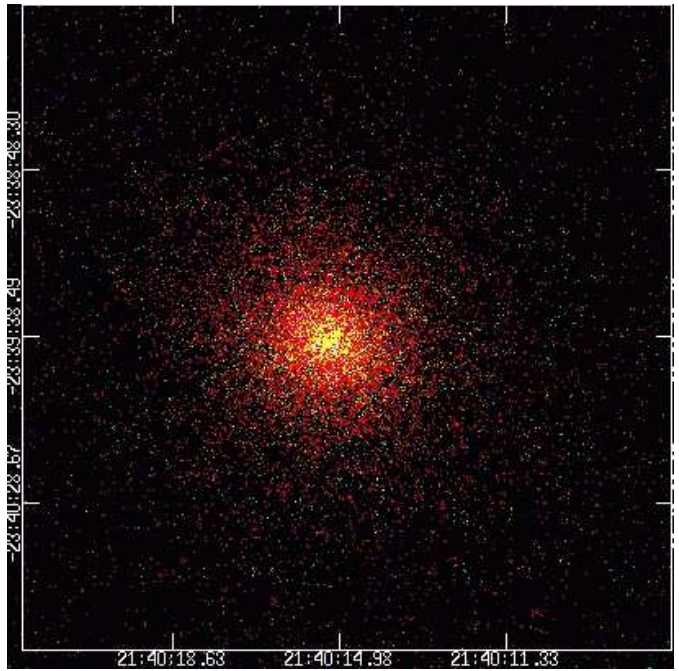,width=0.5\textwidth}  
 \caption[]{True colour image of MS2137; red, green and blue colours correspond to [0.5--2.0], [2.0--4.0], [4.0--8.0] keV bands, respectively. The size of the field of view is $200'' \times 200''$.}
 \label{fig:col_x1}
\end{figure}
We performed our  X-ray analysis on two datasets retrieved from the \textit{Chandra} archive (see Table \ref{tab:obs} for observation log); both the observations were  telemetered in Very Faint mode, and were reduced accordingly. Another \textit{Chandra} dataset is currently available (ID 4974, start date: 2003-11-13, total exposure time:  58.0 ks, PI: Allen). The background light curve of this observation presents  pronounced flares and a  variable quiescent rate. A careful background light curve screening could solve the flaring  issue, allowing to disentangle the ICM emission from the spurious signal. However, unlike the observations we used for our analysis,  Obs.4974 presents a  widely-scattered and variable quiescent count rate. This  feature makes it hard to distinguish  the cluster emission  from possible spurious events, and could affect the background estimate. Moreover, the final net exposure time for this dataset was only $\simeq 16$ ks. So, to prevent systematic errors  associated with an incorrect background estimate, we decided to discard this dataset.
 
\begin{table}
\caption{\textit{Chandra} observations summary.}
\label{tab:obs}
\begin{tabular}{cccccc}
\hline
Obs.ID & Start Time & Total         & Net           & PI\\
       &            & Expos. [ks]  & Expos.[ks]  & name\\
\hline
928    &   1999-11-18 08:37:37  & 44.17 & 20.7      & Wise \\
5250   &   2003-11-18 20:06:43  & 41.09 & 25.0      & Allen\\
\hline		      
\end{tabular}
\end{table}

 The data were reduced using the \textit{CIAO} data analysis package (version 3.4) and the calibration database CALDB 3.4.2; here we summarize briefly the reduction procedure\footnote{See the \textit{CIAO} analysis guides for the data reduction:\\ $\rm cxc.harvard.edu/ciao/guides/$.}. \\ We reprocessed the level-1 event files  to include the appropriate gain maps and calibration products and to reduce the ACIS quiescent background\footnote{For a complete discussion on this topic, see \\ $\rm cxc.harvard.edu/cal/Acis/Cal\_prods/vfbkgrnd/index.html$.}. We used the \texttt{acis\_process\_events} tool to check for the presence of cosmic-ray background events, correct for spatial gain variations due to charge transfer inefficiency and re-compute the event grades. We selected the events flagged with grades $0,2,3,4$ and filtered for the Good Time Intervals associated with the observations. We  identified bright point sources using the script \texttt{vtpdetect}; the results were subsequently checked through visual inspection. We masked out the point sources and corrected for the discarded  area.  \\ 
A careful screening of the background light curve is necessary to identify background flares  \citep{markevitch2003}.
The background light curve  was extracted with a  time bin size of $\simeq$ 1 ks in the energy range [2.5--7] keV, which is the most sensitive to common flares for the S3 BI chip, where the cluster was centered in both observations.  We applied the script  \texttt{analyze\_ltcrv} to include only the time periods  inside  the 3$\sigma$ range. We compared the S3 background light curve with the one extracted in the S1 chips using the energy range [2.5--6] keV, to check for undectected faint flares, that were not identified.  The S3 background light curve was examined using the \textit{ChIPS} facilities to identify and exclude further flaring events; finally we selected only time bins within a factor 1.2 of the  apparent quiescent rate, through the script \texttt{lc\_clean}. At the end of the light curve screening, almost half of the initial exposure time was discarded (see Tab. \ref{tab:obs}).

\label{sec:reduc}

\subsection{X-ray analysis}

Cluster mass measurements derived under the hydrostatic equilibrium hypothesis are strongly dependent on the  temperature profile, as demonstrated by \citet{rasia2006}; therefore, any  temperature bias could  affect the mass estimates. 
We extracted the background spectrum from the blank-field background data sets provided by the ACIS calibration team,  to derive ARF and RESPONSE matrices consistent both with the source and  the background spectrum. The blank-sky observations underwent a  reduction procedure comparable to the one applied to the cluster data, after being reprojected onto the sky according to the observation aspect information.  We derived the background re-normalization factor in the  [9.5--11.5] keV band for the Observation 5250 and in the [8.0--9.8] keV band for the  Observation 928. The normalization factor was  derived in different bands for the two observations because of a difference in the energy upper  limit of the FEF (FITS Embedded Function) calibration files associated to the observations (for the latter observation, the upper limit  is $\simeq  9.8$ kev). Currently, the energy range defined for the spectrum extraction  must match the one in the FEF files  to derive a RESPONSE matrix using the   \textit{CIAO}  package tools\footnote{See cxc.harvard.edu/ciao/bugs/mkwarf.html.}.  However, the observed flux in both these bands can be safely attributed to the background component, so the difference in the  blank-sky  renormalization bands has a negligible impact in the subsequent analysis.\\
The background in the soft band is variable both in time and in space, so we verified whether the soft X-ray background  derived by the  blank-sky datasets is consistent with the observed  one. If they were significantly different, one should take into account this  factor in the X-ray spectral fitting. For both datasets we extracted a spectrum in a region free of cluster emission, to which we subtracted a spectrum derived in the same region of the blank-field dataset. We fitted the residuals in the [0.4--1] keV energy band with a MEKAL model, without an absorption component and broadening the normalization fitting range to negative values. The results indicate that a  corrective component is not necessary. 

\subsection{Spectral fitting}
\label{sec:spec_fit}
\begin{figure}
\psfig{figure=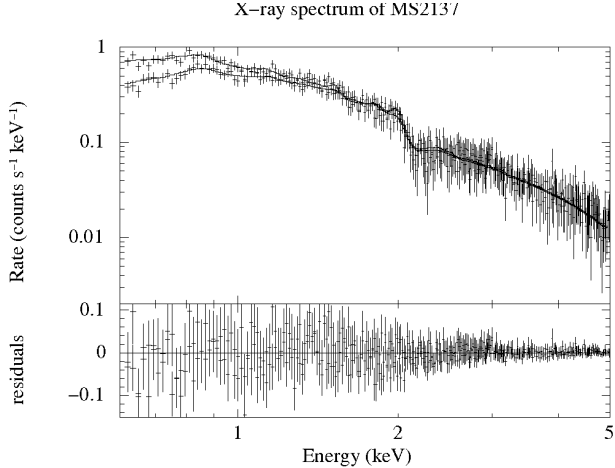,width=0.5\textwidth,angle=-90} 
\caption[]{(Top panel) The X-ray spectrum of MS2137, obtained combining Obs. 928 and Obs. 5250. The spectrum was reduced from an aperture of $2.45$ arcmin, and shows  the energy range $[0.6-5.0]$ keV. We  fixed the Galactic absorption to the value reported in  \cite{dickey1990}. The normalization of the model was left as a free parameter for both datasets. The plotted spectrum was derived with a grouping of  20 counts per bin. The best fit for a  \texttt{MEKAL} model with untied normalizations is shown as a solid line. (Bottom panel) Fit residuals. } 
 \label{fig:spec_x}
\end{figure}

  We derived the cluster X-ray flux image  through the combined, point-source removed images in the energy range [0.3-8.0] keV, properly dividing by the exposure maps for the two observations.  
  The spectra were extracted on concentric, circular  annuli centred  on the X-ray surface brightness centroid, sized to contain at least 5000 net source counts, up to a net count rate of $\simeq 40\%$. We selected 7 annuli, up to 2.45 arcmin.  The X-ray emission appears well centred on the brightest cluster galaxy (BCG). The shift between the estimated X-ray centroid and the BCG center is $\simeq 1$ arcsec: the uncertainty on the X-ray centroid estimate is comparable to the applied smoothing scale ($1.5$ arcsec).   We used the \textit{CIAO} \texttt{specextract} tool both to extract the source and background spectra and to construct ancillary-response and response matrices. The spectra were fitted  in the [0.6--7.0] keV range, except for the last annulus, where, due to the higher background level, we restricted the analysis to the range [0.6--5.0] keV.
   We used the \emph{XSPEC} software package  \citep{arnaud1996} to analyse the spectra, that were fitted adopting an optically-thin plasma emission model (the \texttt{MEKAL} model; \citealp{mewe1985,kaastra1993}), with an absorption component, described by the T\"ubingen--Boulder model (\texttt{tbabs} model; \citealp{wilms2000}), to take into account the effect of Galactic absorption on the observed flux.  \\
The Galactic absorption was fixed to the value inferred from radio HI maps in \cite{dickey1990}, i.e. $3.55\times 10^{20}$ cm$^{-2}$, with the cluster-averaged  values of Galactic absorption derived from the X-ray spectra fit being consistent with the radio value $(3.9\pm 0.8 \times 10^{20}\rm  cm^{-2}$ and $4.2\pm 0.9\times 10^{20}\rm cm^{-2}$ for the Observations 928 and 5250, respectively).  Thus the free parameters in our model were the temperature, the metallicity and the normalization of the thermal spectrum. \\
The spectral data were not rebinned to keep a high spectral resolution: we only grouped the channels to have at least one count in each channel. As a consequence, we applied the \textit{c-stat}  statistic to the spectral fitting. \\
The two observations were  at first analysed individually, to assess the consistency of the datasets and to exclude any systematic effects that could influence the combined analysis. A significant difference was detected in the spectral fit results for the most distant radial bin, supposedly for  the higher background in Obs.5250. Thus the spectrum relative to Obs.5250 was discarded. We proceeded with the joint analysis of the two datasets, tying the temperature and metallicity parameters and leaving the normalization unlinked. The fit results from the spectral analysis of the two combined datasets are listed in Tab. \ref{tab:tab_xspec}. The resulting metal abundance profile  presents a central peak with a smooth external decline,  as is generally observed for dynamically relaxed clusters. Moreover, the  temperature decrease in the central region associated with the cluster cool core is evident in Fig. \ref{fig:spec_x}. Several earlier studies stated that the presence of a cool core, if neglected,  could be a possible source of error in the X-ray mass estimate, resulting in an underestimate of the cluster virial temperature and mass  \citep{allen1998,hallman2006,rasia2006}.  On the contrary, if the gas temperature decrement is properly taken into account, cool core clusters should be  ideal targets for mass analyses, being typically dynamically relaxed systems. Since our X-ray mass estimate relies directly on the \emph{measured} gas temperature profile, the temperature decrement in the center is not an issue in the mass derivation.

\begin{table}
\caption{X-ray spectral fit results. The analysis was performed in the [0.6--7.0] keV energy range except for the last annulus, where we considered the range [0.6--5.0] keV. The columns list the outer radius of the annulus where the spectral extraction was performed, the best-fit values obtained for the  free parameters of the model (i.e. the intra-cluster gas temperature, the normalization of the X-ray spectrum and the metallicity) and the $c$-statistic result (the degrees of freedom for each annulus are in square brackets). }
\begin{threeparttable}
 \input{specres.tex}

 \begin{tablenotes}[para] $^{a}$ The results obtained for the last annulus refer to a fit performed for Obs. 928 only. 
   \end{tablenotes}
        \end{threeparttable}
        \label{tab:tab_xspec}
\end{table}

\subsection{X-ray mass profile}
\begin{figure}
\psfig{figure=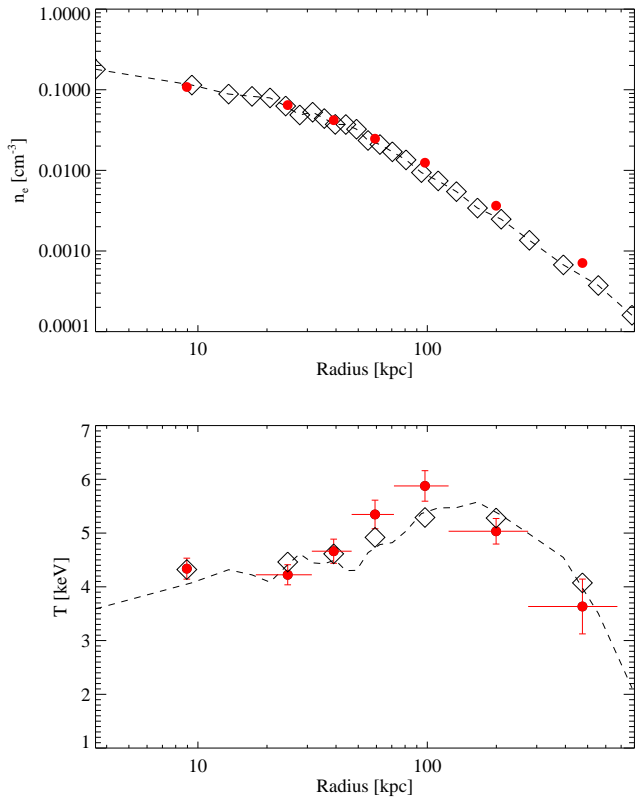,width=0.5\textwidth}
 \caption{(\emph{Top panel}) Deprojected gas density profile. Red circles mark the values determined by the normalization of the thermal spectra, diamonds represent the gas density values obtained combining both spectral and  surface brightness measurements; the dashed line shows the best-fit profile. (\emph{Bottom panel})  ICM temperature profile. Here we show the observed temperature profile  (red circles, $T_{\rm obs}$ in the text), directly inferred from the spectral analysis, the deprojected temperature values  (diamonds, $T_{\rm pred}$ in the text),  rebinned to the same intervals of the observed profile,  and the deprojected best-fit profile (dashed line).The small scale fluctuations of the temperature best-fit profile reflect the local variations in the gas density, since  the condition of hydrostatic equilibrium is imposed between the smooth total mass profile and the gas pressure  profile $P_{\rm gas} \propto n_{\rm gas}\times T_{\rm gas}$ (see Eq. \ref{eq:mtot} in the text).}
 \label{fig:prof1}
\end{figure}

In this work we recover the deprojected profiles of gas temperature and density  without any parametrization of the ICM properties, which could introduce considerable systematic errors in the mass measurement results (\citealp{rasia2006}). Our estimate relies on the assumptions of spherical symmetry and  hydrostatic equilibrium and on the choice of an analytic  model for the total mass.   In the case of MS2137  these assumptions seem to be reasonable, since, e.g., the strongly peaked central surface brightness and a roughly circular X-ray emission  can generally suggest a relaxed dynamical state. \\
Here we summarize the most relevant aspects of the technique we have applied; for full details, see \citet*{ettori2002,morandi2007a,morandi2007b}.\\
From the spectral analysis we derived for each annulus  the projected gas temperature $T_{\rm ring}$, metal abundance $Z_{\rm ring}$ and the normalization $K$ of the model; the latter provides an estimate for the Emission Integral $EI = \int \rm n_{\rm e} n_{\rm p} dV \simeq   0.82 \int \rm n^{2}_{\rm e}\, dV$. 
To derive the value of these quantities in volume shells, one has to calculate the  intersection of each volume shell  with 2-dimensional annuli.   Geometrical considerations \citep*{kriss1983,buote2000} allow to evaluate the intersections through an upper triangular matrix (\textit{V}) where the last pivot represents the outermost annulus,  under the assumption of spherical symmetry. In this way we can recover the deprojected values applying some array operations. \\
To calculate the deprojected density profile, we  combined the spectral information and  the cluster X-ray surface brightness (hereafter SB) profile. The  SB provides an estimate of the  volume-counts emissivity $F_{\rm bin} \propto \rm n_{\rm e\ \rm bin}^{2}T(\rm bin)^{1/2}/D^{2}_{\rm L}$. By comparing this observed profile  to the values predicted  by a thermal plasma model (with  temperature and metallicity equal to the measured quantities and taking into account the  effect of Galactic absorption) one can solve for the electron density. In this way, we  obtain  a gas density profile which is better resolved than a spectral--only profile (see the top panel of Fig. \ref{fig:prof1}). We computed the SB in 35 circular annuli. They were derived from the X-ray  flux  image in the energy range [0.5-2.0] keV, thus similarly to the spectral extraction annuli (see $\S\ $\ref{sec:spec_fit}), but applying a binning of 1000 counts. The X-ray centroid is the same for both profiles. We rederived the mass estimate  calculating the SB profile over the range $[0.3 -  8.0]$ keV, to  verify if the energy range choice  would affect our results.  \\
 The total  mass was then constrained as follows. The observed deprojected temperature profile in volume shells $T_{\rm shell}$ was obtained through the relations:
\begin{eqnarray}
n_{\rm e} &=& \left[ ({\bf Vol}^T)^{-1} \# (EI/0.82) \right]^{1/2}, \\  
 \epsilon &= &({\bf Vol}^T)^{-1} \# L_{\rm ring},\\
\epsilon T_{\rm shell}& = &({\bf Vol}^T)^{-1} \# (L_{\rm ring} T_{\rm ring}), \\
 \epsilon Z_{\rm shell}& =& ({\bf Vol}^T)^{-1} \# (L_{\rm ring} Z_{\rm ring}),   
\end{eqnarray}  
where  the symbol $\#$ indicates a matrix product. \\
We compare it to the predicted temperature profile  $T_{\rm pred}$, obtained by inverting the equation of hydrostatic equilibrium between the dark matter potential and the potential of the intracluster plasma, i.e.:
\begin{eqnarray}
-G \mu m_{\rm p} \frac{n_{\rm e} M_{\rm tot, model}(<r)}{r^2} =
\frac{d\left(n_{\rm e} \times kT_{\rm pred}\right)}{dr};
\label{eq:mtot}
\end{eqnarray}
where $n_{\rm e}$ is the deprojected electron density. Our best-fit mass model was determined  by comparing the observed $T_{\,\rm obs}$ profile to the predicted one, which depends on the mass model parameters.  So we minimized:
\begin{equation} 
\chi^2 = \sum_{\,\rm \,rings}\,\left(
\frac{T_{\,\rm obs} - T_{\,\rm pred}}{\sigma_{\,\rm obs}} \right)^2.
\label{eq:chi2x}
\end{equation} 
The $T_{\rm pred}$ profile was rebinned to the spectral scale.
In the bottom panel of Fig. \ref{fig:prof1} we show the projected temperature profile as determined by the spectral analysis, the deprojected best-fit profile and the deprojected, rebinned  profile. 

\begin{figure}
\psfig{figure=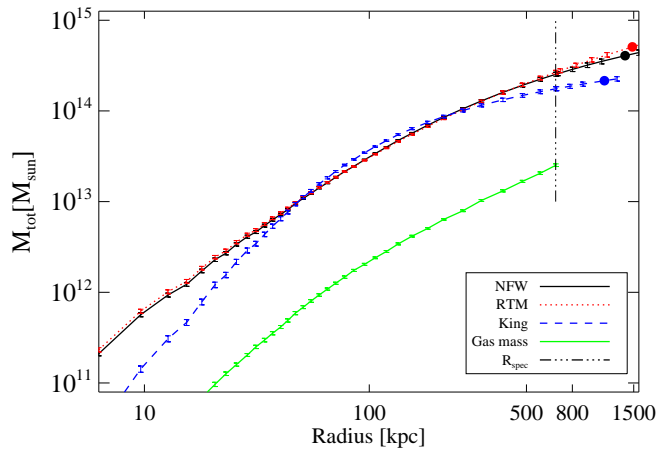,width=0.5\textwidth} 
\caption{MS2137 total mass profiles, as determined through the X-ray analysis for different choices of parametrised mass models. The vertical line indicates the boundary of the spectral extraction,  the filled circles mark the value of $\rm R_{200}$ for the corresponding mass profile. The measured gas mass profile  is also shown as a green solid line.}
 \label{fig:mass_x3}
\end{figure}

\begin{figure}
\psfig{figure=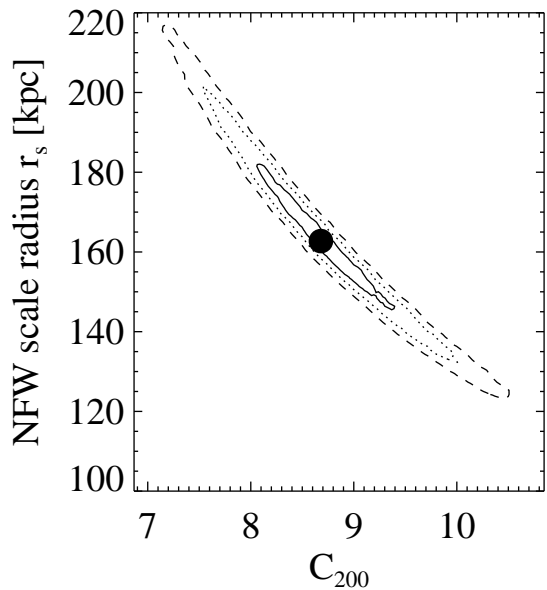,width=0.5\textwidth}
 \caption{Confidence levels determined through the X-ray analysis for the NFW profile parameters $\rm c_{200}$ and $\rm r_{s}$. The contours indicate the $1\sigma,2\sigma$ and $3\sigma$ c.l., and  the black circle marks the best-fit values.}
 \label{fig:levx}
\end{figure}

We assumed an NFW density profile, which  can be expressed as:
 \begin{equation}
\frac{\rho_{\rm NFW}(r)}{\rho_{\rm c}}=\frac{\delta_{c}}{(\rm r/\rm r_{\rm s})(1+\rm r/\rm r_{\rm s})^{2}},
\label{nfwprofile}
\end{equation}
\noindent where $\rm r_{\rm s}$ is the scale radius of the halo, $\delta_{c}$ is its characteristic overdensity and  $\rho_{\rm c}$ is the critical density of the universe. We will  define the NFW profile using the scale radius and  the concentration parameter $\rm c_{200}$, i.e. the ratio of the radius where the halo matter density is 200 times the critical density ($\rm r_{200}$) to the scale radius: $c_{200}=\rm r_{200}/\rm r_{\rm s}$. \\
 The best-fit values are determined by minimizing the statistical estimator defined in Eq. \ref{eq:chi2x}.       
 A first minimization  is performed in the ranges $0.5\!\!\leqslant \!\rm c_{200}\!\leqslant 20.0$ and $10\ \rm kpc\!\!\leqslant \!r_{\rm s}\! \leqslant \! \rm R_{\rm spat}$ (where $\rm R_{\rm spat}\! \simeq 676\ \rm kpc$ is the outer radius of the surface brightness profile). In this way a first estimate of the best-fit values is obtained. A second minimization  is then performed in the $X_{i}\pm 3\sigma$ range, being $X_{i}$ the mean values inferred through the first minimization. \\
The mass model parameters obtained through our X-ray analysis,  deriving the SB profile in the  energy range $[0.5 -  2.0]$ keV, are:
\begin{tabular}{c@{\@{\hspace{10pt}}}c@{\@{\hspace{10pt}}}c@{\@{\hspace{10pt}}}c@{\@{\hspace{10pt}}}c}
\hline 
\multicolumn{5}{c}{{\textsc{Best fit X-ray results}}} \\
\hline
$\rm r_{\rm s}$ & $\rm c_{200}$ & $\rm M_{200}$ &  $\rm R_{200}$ & $\chi^{2}_{\rm red}$[d.o.f.]\\
 $\rm[kpc]$ & & $[10^{14}\rm M_{\odot}]$ & $\rm[Mpc]$ &\\ 
\hline
\hline
$162.72^{+18.6}_{-16.3}$  &$8.68^{+0.7}_{-0.9}$  & $4.4 \pm 0.3$  & $1.41\pm 0.03$  & $2.1[5]$\\
\hline
\end{tabular}  \\
 The  confidence levels for the NFW  parameters are shown in Fig. \ref{fig:levx}. \\
 The choice of the energy range over which the surface brightness is estimated does not significantly  affect our results: for example, deriving the SB profile in the wider range $[0.3 -  8.0]$ keV we obtain the following  values:
 \begin{enumerate}
\item  $\rm r_{s} =  153.8^{+17.2}_{-20.7}\ \rm kpc$,
\item  $\rm c = 8.93^{+1.0}_{-0.6}$,
\item  $\rm M_{200} = 4.1 \pm 0.3 \times10^{14} \rm \ M_{\odot}$,
\item $\chi^{2}_{\rm red}\rm[d.o.f.] = 3.1[5].$
\end{enumerate}
The  gas mass profile and the total mass profile are shown in Fig.\ref{fig:mass_x3}. To indicate how much the X-ray mass profile  depends on the choice of the  density profile, in Fig.\ref{fig:mass_x3} we show also  the mass profiles derived assuming two other parametric models, i.e. the Rasia-Tormen-Moscardini (hereafter RTM) profile \citep*{rasia2004} and the King profile \citep*{king1962}. 
The RTM model was deduced  from the direct  analysis of non-radiative $N$-body hydro-simulations, without any assumption on the gas  equation of state or on the cluster dynamical equilibrium. The resulting DM density distribution is:
 \begin{equation}
 \label{eq:rtm1}
\tilde{\rho_{R}}(r) = \frac{\rho_{0,R}}{r(r+r_{\rm \star})^{1.5}}.
\end{equation}
 The King profile \citep{king1966,rood1972} was derived under the assumption of an isothermal particle distribution, and can be expressed as:
 \begin{equation}
 \label{eq:king}
{\rho_{\rm K}(r)} = \frac{\rho_{\rm 0,K}}{(1+(r/r_{\rm c})^{2})^{3/2}}.
\end{equation}

The mass estimates which were obtained adopting the RTM and the King  mass model  are  $\rm M_{200, \rm RTM}=5.1\pm 0.3 \times 10^{14}\rm M_{\odot}$ and $\rm M_{200, \rm King}= 2.2 \pm 0.2 \times 10^{14}\rm M_{\odot}$. The  corresponding $\chi^{2}_{\rm red}$ values are $3.7$ and  $4.1$ (over 5 degrees of freedom), respectively. When assuming the RTM or the King density profile, the surface brightness was estimated in the energy range $[0.3 -   8.0]$ keV. The  best-fitting model  is the NFW profile, even though the RTM profile is not  a significantly worse fit. Clearly, the extrapolated mass derived through a profile shallower in the external regions is lower than the mass estimates obtained with  more cuspy parametrizations.
 
We have compared our X-ray estimate with the value predicted by the $M\!-\!T_{X}$ relation.  For instance, \citet*{arnaud2005}, through the analysis of 10 nearby, relaxed galaxy clusters observed with \emph{XMM-Newton}, deduced  the following relations  for clusters with  $kT \geq 3.5\ \rm keV$:
 \begin{eqnarray*}
  h(z)M_{200}&=&5.74 \times \frac{kT}{5\ \rm keV}^{1.49}, \\
  h(z)M_{500}&=&4.10 \times \frac{kT}{5\ \rm keV}^{1.49},
 \end{eqnarray*} 
where $h(z)$ is the Hubble constant normalised to its local value and $kT$ is the spectroscopic temperature over the  $[0.1  -   0.5]\ \rm R_{200}$ region (corresponding to $[0.15  -   0.75]\ \rm R_{500}$)\footnote{Accurate predictions on the X-ray scaling relations are also presented  in \cite*{nagai2007}, through the results obtained by   \cite{vikhlinin2006} analysing \textit{Chandra} observations of 13 low redshift, relaxed clusters. The scaling relations in \cite{nagai2007} are calibrated with the gas temperature derived in the wider radial range $70$ kpc $< \rm r <\rm R_{500}$. We applied the scaling relations found by \citet{arnaud2005}  to refer  to a spatial range where our spectroscopic temperature estimate is more robust. However, \citet{arnaud2005} asserted that their  scaling relation estimates are very similar to those derived in  \cite{nagai2007}, so our conclusions would be unchanged.}.  The authors derived the temperature $T_{X}$  from a single-temperature fit to the integrated spectrum. The highest redshift cluster in the sample is at $\rm z \simeq 0.15$. We  estimated the value of $kT$ in a similar manner to  \citet{arnaud2005}: we  extracted the X-ray spectrum in the energy interval $[0.6  -   5]$ keV and in the same radial range  $[0.1 - 0.5] \rm R_{200}$. We derived the value of  $\rm R_{200}\!=\!1.37\pm 0.04$ Mpc  fitting the data with an NFW profile for a density contrast $\delta\!=\!200$. We considered only the Obs. 928, due to the higher background in the external regions for Obs. 5250 (see $\S\ $\ref{sec:spec_fit}).   The spectroscopic gas temperature over the spatial range  $[0.1  -   0.5]\ \rm R_{200}$  is $T_{X} = 4.1\pm0.3$ keV (the  $c-$statistic value for the spectral fit is 293.2 over 296 degrees of freedom). This value leads to  $\rm M_{200,\rm scal} \simeq 3.7\times 10^{14}\rm M_{\odot}$ and $\rm M_{500,\rm scal} \simeq 2.6\times 10^{14}\rm M_{\odot}$. These estimates are  lower than ours (i.e. $\rm M_{200} \simeq 4.4\pm 0.3\times 10^{14}\rm M_{\odot}$ and $  \rm M_{500} \simeq 3.5\pm0.2\times 10^{14}\rm M_{\odot}$), but are compatible at a $4\sigma$ level with our results. The X-ray mass of  MS2137 seems to be consistent with the $M\!-\!T_{X}$ scaling relation found for relaxed galaxy clusters.
 
\section{Strong lensing analysis}
\label{sec:sl}

\subsection{Overview}
\begin{figure}
\psfig{figure=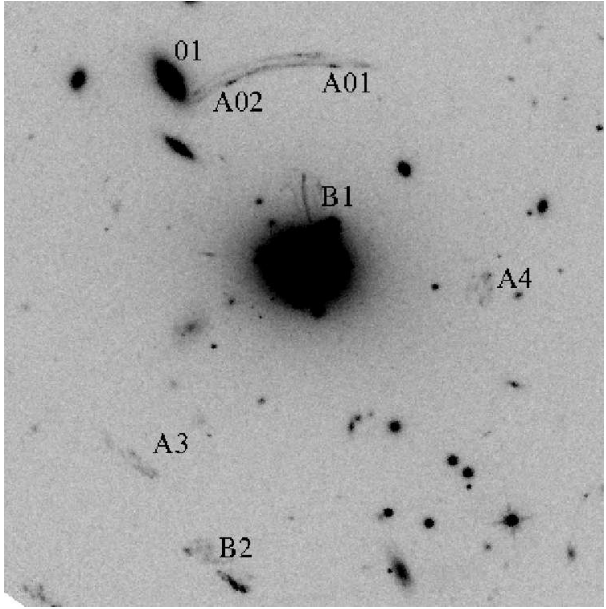,width=0.45\textwidth} 
 \caption[]{Multiple image interpretation of the lens system, overplotted on the HST/WFPC2 image of MS2137 (filter: F702W; number of combined images: 10; total integration time: 22.2 ks; proposal ID: 5402; PI: I.Gioia). The size of the field of view is $\simeq 42\ \rm arcsec \times 45\ \rm arcsec$; the coordinates of the images used as constraints in the strong lensing analysis are listed in Table \ref{tab:tab_ima}.}
 \label{fig:ms1}
\end{figure}

MS2137.3--2353 is a well-known lensing system;  the radial feature in its double-arc configuration is the first  ever detected \citep{fort1992}. The radial arc may be very useful in constraining the cluster central density slope. For axially simmetric lenses, the radial and tangential critical curves  arise if $\lambda_{\rm r} = 1 - (d/dx)(m/x)$ and $\lambda_{\rm t}= 1 - m/x^{2}$ vanish, respectively, $\lambda_{\rm r}$ and $\lambda_{\rm t}$ being  the eigenvalues of the Jacobian matrix of the lens mapping (see   \citealt{bartelmann1995} for an extended review). Thus, the tangential and radial arc positions allow one to constrain the enclosed total  mass and the mass distribution derivative, respectively. \\
We carried out a strong lensing analysis on deep HST data retrieved from the Space Telescope Archive\footnote{archive.stsci.edu/}.  The data consists of a WFPC2 association image, taken with the F702W filter and obtained combining 10 exposures for a total integration time of $22.2$ ks\footnote{Proposal ID: 5402; PI: I.Gioia.} (a portion of the HST image is shown in Fig. \ref{fig:ms1}). \\
The  tangential and radial arc distances from the BCG center are  $\simeq 15$ and $\simeq 5$ arcsec: the former is composed of two images with reverse parity (A01 and A02 in Fig. \ref{fig:ms1}, following the notation of \citealt{gavazzi2005}), associated with the counter-images A2 and A3. A fifth demagnified image lying near the center of the BCG is predicted if the central density slope is shallower than the isothermal model slope. Thus the position  of the central image would provide an additional constraint on the inner density (for example, see  the results of  \citealt*{gavazzi2003}), and as a consequence on the BCG mass.   However,  the BCG-subtracted image presents  residuals in the central region, where the fifth image, if present, is predicted. The noise level after the BCG subtraction  makes a clear identification of the central image difficult.  For these reasons the central counter-image was not included as an additional constraint. \\
Spectroscopic   measurements \citep*{sand2002} stated that the two groups of images arise from 2 different sources at redshift $z_{\rm tang}=1.501$ and $z_{\rm rad}=1.502$; recent data \citep*{soucail2007} confirmed that the counter-image of the radial arc B1 is the feature B2 (see Fig. \ref{fig:ms1}).\\ 
In the next section we will present  a parametric strong lensing reconstruction of the cluster mass distribution, performed using the \emph{Lenstool}\footnote{The code is available at the url www.oamp.fr/cosmology/lenstool.} analysis software \citep{kneib1996,jullo2007}. \\

\begin{table}
\caption[]{Coordinates of the multiple images considered in the MS2137  strong lensing modeling. Coordinates are  in arcsec with respect to the BCG center. We did not include as constraint the fifth central image, that was not clearly identified. Spectroscopic measurements \citep*{sand2004} indicate a redshift of $\rm z_{tang}=1.501$  and  $\rm z_{rad}=1.502$ for the first and  the second system, respectively.}
\label{tab:tab_ima}
 \input{tab_knotsa.tex}  
\end{table}

\subsection{Lens modeling}
\label{sec:mod}
To identify the multiple image systems,  we started by referring to the configuration provided by \cite{gavazzi2005} (hereafter G05), the most detailed knots catalogue available in the literature. We then iteratively refined the system by selecting several sets of ellipses, likely belonging to a common source area (their coordinates are listed in  Table \ref{tab:tab_ima}). \\
Our lens mass model consists of a cluster--scale mass component and of two galaxy--scale components, namely the BCG and the galaxy near the tangential arc. The latter (flagged $01$ in Fig. \ref{fig:ms1}) was included to account for its likely perturbative effect, due to its proximity to the tangential arc \citep{meneghetti2007b}. 
We modeled the cluster scale component using an elliptical NFW profile, as  implementend in the \emph{Lenstool} code (for a detailed discussion on the lens properties of this density profile, see  \citealp*{bartelmann1996,meneghetti2003a}: for full details on the elliptical NFW implementation in \emph{Lenstool} the reader can refer to \citealt{golse2002}). \\
 Our  main clump model has seven free parameters: the coordinates of the center of mass   ($\rm x_{\rm c},\rm y_{\rm c}$), the ellipticity of the mass distribution ($e$), the position angle ($\theta$), the characteristic velocity dispersion $\sigma_{0}$ and the NFW profile parameters (the concentration $\rm c_{200}$ and the scale radius $\rm r_{\rm s}$). 
The center of mass  was initially set to the BCG centroid, but  it was  allowed to vary by $\pm15$ arcsec; the ellipticity and the position angle of the dark matter clump were also, as a first estimate, set equal to the BCG values, then were optimized in the ranges  $[0.0\leqslant e \leqslant 0.5]$ and $[130.0\leqslant\!\theta\!\leqslant 160.0]$ degrees. The optimization ranges for the scale radius and the concentration parameter are $[50\leqslant \rm r_{\rm s} \leqslant 650]$ kpc and $[0.5\leqslant\!\rm c_{200}\!\leqslant 20.0]$. The choice of these  intervals is motivated by cluster analysis results  from previous studies. Of course, strong lensing allows to constrain only the inner matter distribution, so  the estimate of the virial radius - and therefore of the concentration parameter - using strong lensing alone is  less robust; however, we decided not to fix any mass distribution parameter, but rather to sample them in wide ranges to avoid biasing our results.\\
The galaxies were modeled either  as dual Pseudo Isothermal Elliptical Mass Distribution - hereafter dPIE - with a null core radius,  or as elliptical singular isothermal sphere  - hereafter SIE (see \citealp{natarajan1997}, \citealp*{limousin2005}, \citealp{eliasdottir2007} for the properties of these potentials and  for full details on their implementation in the \emph{Lenstool} code).
Unlike the studies of \citet{gavazzi2005,sand2008}, we did not model the BCG stellar velocity distribution in our analysis. We assumed a  parametric model for the BCG mass (dPIE or SIE), so  its mass is determined by the central velocity dispersion (and the scale/core radius values in the dPIE case).  We are aware that stellar dynamics data would help to solve the existing degeneracy between the stellar mass and  the  mass of DM and gas  in the cluster center. However, here we are more interested in determining the shape of the total mass profile at radii larger than those where the stellar mass is a non negligible mass component.

\begin{figure}
\psfig{figure=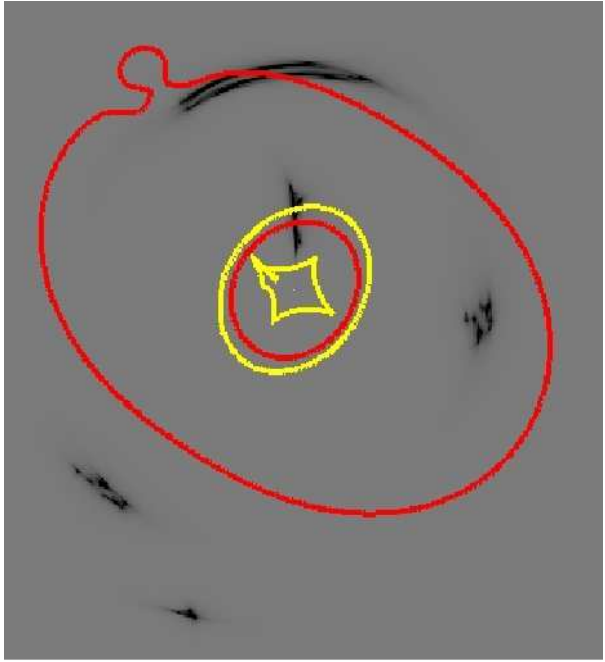,width=0.45\textwidth} 
\caption[]{Images predicted by our best-fit lens model. In this case the galaxies included in the fit (i.e. the BCG and the galaxy flagged $01$ in Fig. \ref{fig:ms1}) were modeled as dPIE  potentials. The size of the image is $\simeq 42\ \rm arcsec \times 45\ \rm arcsec$. The red [yellow] lines represent the critical [caustic] lines.}
 \label{fig:predima}
\end{figure}

We fixed the galaxies ellipticity and orientation  to the  values  derived by the HST image, using the IRAF external package \texttt{isophote} of the \texttt{stsdas} distribution. When modeling the galaxies as dPIE mass distribution, the values we imposed to the dPIE scale radius are almost the same of  \cite*{sand2008}, so $\rm r_{\rm cut,BCG}$= 22 kpc and $\rm r_{\rm cut,gal}$= 5 kpc. The other free parameter in the galaxy optimization was the central velocity dispersion $\sigma_0$; the spectroscopic measurements obtained by \citet*{sand2002} with   the Echelle Spectrograph and Imager on the Keck II telescope indicate a central value for the BCG stellar velocity of  $\sigma_{\rm measured,BCG}\simeq 325\ \pm 35\ \rm km\ s^{-1}$. However, the cD measured  velocity dispersion  cannot be directly related to the ``characteristic'' one in galaxy mass models, since the observed dispersion depends upon the unknown dark matter potential contribution to the stellar dynamics. The uncertainty on the velocity dispersion parameter could potentially introduce additional errors in the strong lensing modelling \citep{wu1993}. Despite of this,  the BCG stellar mass is an important lens component  (\citealp*{meneghetti2003b}, \citealp{hilbert2008}), especially when modeling radial arcs, which, due to their position very close to the cluster center, are extremely sensitive to the  cuspy distribution of luminous matter (see the earlier findings of \citealp{miralda1995} on this topic).  We thus imposed  the measured  $\sigma_{\rm measured,BCG}$ as an upper limit for the velocity dispersion optimization range, and set as  lower  limit the  value of $\sigma_{\rm 0,BCG}=230\rm \ km\ s^{-1}$, that is plausible given the  high luminosity of the BCG. A discussion on the effect of this choice on the lensing results is presented in $\S\ $\ref{sec:sl_res}.\\
The optimization procedure  of the \emph{Lenstool} code is performed in the source plane, by mapping the positions of the input multiple images  back to the source plane, and requiring they have a minimal scatter \citep{jullo2007}.
 The goodness-of-fit is evaluated through the following $\chi^2$ estimator: 
 \begin{equation}
 \label{eq:chi2}
 \chi^2{=}\sum_j  \chi^2_{\mathrm{pos}} (j),
 \end{equation}
  where the subscript $j$ refers to the multiply imaged systems.

We assumed  the  uncertainty in the lensed image positions to be $\sigma_{I}=0.3''$: mapping it back to the source plane one obtains an estimate of the error in the source plane. 

\subsection{Strong lensing results}
\label{sec:sl_res}
The optimization and the error measurement are performed through the Bayesian optimization method  implemented in \emph{Lenstool}, imposing an optimization rate of  $\delta \lambda =0.1$. \\
A summary of the best-fit values for  two lens configurations  is reported in Table \ref{tab:lens_comp}.
%...
\begin{figure}
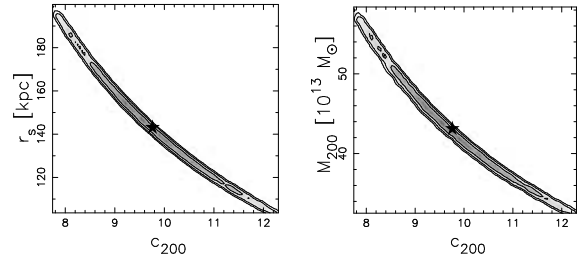

\begin{tabular}{cc}
\psfig{figure=fig8.ps,width=0.2\textwidth} &
\psfig{figure=fig9.ps,width=0.2\textwidth} \\
 \end{tabular}
\caption[]{Confidence levels for  the NFW mass model parameters  determined through a strong lensing analysis. The contours show the parameter  non-marginalized probability density, and refer to the model $[1]$ in Table \ref{tab:lens_comp}. [\emph{Left panel}] NFW concentration parameter $\rm c_{200}$ vs the scale radius $\rm r_{\rm s}$; [\emph{right panel}] the concentration parameter  vs the cluster total mass $\rm M_{200}$, extrapolated up to $\rm R_{200}$.}
 \label{fig:cont_lens1}
\end{figure}

In Table \ref{tab:best_pot} we list  the galaxy model values inferred through the strong lensing optimization. The configuration we will take as a reference  in the following is labelled $[1]$. \\ 
The lensed images predicted  by the reference  model are shown in Fig.\ref{fig:predima} and the corresponding likelihood contour levels for the NFW parameters  are shown in Fig.\ref{fig:cont_lens1}. In Table \ref{tab:chi2l} are reported the $\chi^2$ and the mean scatter for each image system. 

 \begin{table*}  
\begin{center} 
\caption[]{Best fit  parameters, obtained for different lens models.  Hereafter, we will refer to the first model as the reference model. Quoted errors correspond to $1\sigma$ confidence level. The columns indicate the lens configuration (the cluster halo model + the galaxy model), the coordinates of the center of mass, expressed in arcsec with respect to the BCG centroid, the  ellipticity and position angle (WCS aligned) of the cluster clump, the scale radius and the concentration parameter (both referred to a halo overdensity of 200),  the extrapolated mass up to $\rm R_{200}$, the characteristic velocity dispersion of the NFW profile implementation in \emph{Lenstool} and  the total $\chi^{2}$, calculated over 15 knots systems, for a total of 50 image areas.  When the  best-fit value is significantly different from the mode of the distribution, the former is reported enclosed in brackets.
The  errors on $\rm x_{c},\rm y_{c}$ and on the ellipticity are $\simeq 0.1$ and $\simeq 0.01$, respectively. Ellipticity \emph{e} here is expressed as  $(a^{2}-b^{2})/(a^{2}+b^{2})$. }
\label{tab:lens_comp}  
\input{tab_comp.tex}  
\end{center}
\end{table*}  

The curvature of the  radial arc, peculiar of this strong lensing system, is not well reproduced by our lens models. Due to its position very close to the cluster center, the radial arc is affected by  the BCG mass and dynamics. Thus  a possible explanation of the offset between the predicted and  the observed radial arc curvature could be, e.g.,  the  effect of a potential with non-constant ellipticity or position angle, that cannot be taken into account with the current mass model implementations  \citep[see][]{gavazzi2003}. Other features of the arc (like its position or its radial extension) are quite well reproduced by our models. So, despite of the uncertainty on its curved shape, the radial arc provides indeed a tight constraint on the position of the radial critical line, and as a consequence on the total mass derivative in the cluster central region.
 \begin{table}  
\caption[]{Best fit parameters for  the galaxies included in the lens system. The galaxies were parameterized as dPIE or SIE potentials in the lens model [1] and [2], respectively (in Tab. \ref{tab:lens_comp} we list the fit results for the other parameters of the lens model). We report the mode of the distributions, with the $68\%$ statistical errors, and the  best-fit values enclosed in round brackets. Values in square brackets were not optimized.}.
\label{tab:best_pot}  
\input{best_pot.tex}  
\end{table}  

\begin{table}  
\caption[]{The results of the source plane optimization performed with the \emph{Lenstool} code for our reference model, labelled [1] in Table \ref{tab:lens_comp}. The columns list the knots system, the $\chi^2$ of the  system, the root-mean-square  computed in the source plane ($\rm rms_s$) and the corresponding one in the image plane ($\rm rms_i$). The last row reports the total $\chi^2$ estimator, referring to a model with 61 degrees of freedom.}
\label{tab:chi2l}  
\begin{center} 
\input{tab_chi2.tex}

\end{center} 
\end{table}  
 
	Resuming the strong lensing results, the best-fit models indicate a negligible shift of the cluster center with respect to the BCG centroid, and a slight offset between the  BCG and the cluster halo projected orientation, $\Delta \theta \cong 5^{\circ}$,  which is lower than the $\Delta \theta \cong 13 ^{\circ}$ found by G05 and by \cite{sand2008}. Such a misalignment could have either a geometrical explanation (e.g. the projection effect of a prolate halo -see G05- or the difference of halos triaxialities \citep*{keeton1997}) or a dynamical origin, linked to the BCG formation process (\citealp*{quadri2003}; see also \citealp{minor2007}). \\
To estimate the magnitude of the systematic error on the lensing mass introduced by the uncertainty on the velocity dispersion, we optimized again  the lens model imposing as lower limit  $\sigma_{\rm 0,BCG}=325\rm \ km/s$. Under these conditions we derived the following values:
\begin{tabular}{@{\@{\hspace{20pt}}}c@{\@{\hspace{15pt}}}c@{\@{\hspace{15pt}}}c@{\@{\hspace{15pt}}}@{\@{\hspace{15pt}}}c}
\hline
\hline
Parameter & Unit & Mode  & Best fit \\
& & &  \\
\hline
 $\rm x_{ c} $ & arcsec &$0.2\pm0.1$\\
  $\rm y_{c}$ &  arcsec &  $0.2\pm0.1$\\
    $ e$ & & $0.11\pm0.01$ \\
  $\theta$   &  deg  & $146.2^{+0.4}_{-0.4}$ \\
  $\rm r_{\rm s}$& kpc  &  $153.1^{+23.3}_{-19.5}$&$[160.1]$ \\
  $\rm c_{200}$ & &$9.21^{+0.92}_{-0.86}$ &$[8.93]$ \\
  $\rm M_{200}$  & $\rm 10^{14} M_{\odot}$  &$4.40^{+0.6}_{-0.5}$&$ [4.59]$ \\
  $\sigma_{0,\rm halo}$ &  km $\rm s^{-1}$  & $1359.0^{+31.3}_{-25.5} $ \\
    $\sigma_{0,\rm BCG}$ & km $\rm s^{-1}$ &$ 327.0^{+3.2} _{-0.15} $&$[325.4]$\\
    \hline
  $\chi^{2}_{\rm tot} [\rm d.o.f.]$ & & $61.1[61]$\\
  \hline
  \hline
\end{tabular}
	
The values in square brackets indicate the best-fit results if they are significantly different from the mode of the distribution. The parameter values    in this case  are within $2\sigma$ respect to the ones presented in Tab. \ref{tab:lens_comp}. The results  obtained  decreasing the BCG velocity dispersion (and thus its mass) are shown in $\S\ $\ref{sec:prev_comp}. So, considering the results obtained for the two boundary conditions (no halo contribution to the BCG potential well and no mass in the BCG) we can estimate that, regard to the strong lensing mass, the systematic errors  associated with the uncertainty in the BCG mass account for an additional $\simeq 25\%$ in the final error budget. We would like to stress again that this additional error component arises in the case of galaxy clusters with central lensed features, when trying to disentangle the central galaxy mass from the underlying halo mass and \emph{when no additional constraint} (e.g. the BCG stellar dynamics) \emph{is included}. \\
 Anyway the NFW parameters are quite well constrained by the cluster strong lensing features, and we did not detect a high concentration for this cluster, as stated in previous works in the literature (see \citealp{comerford2007} and $\S\ $\ref{sec:prev_comp} for further discussion).

\section{X-ray and strong lensing comparison}
\label{sec:comp}
\begin{figure*}
\begin{center} 
\begin{tabular}{cc}
\psfig{figure=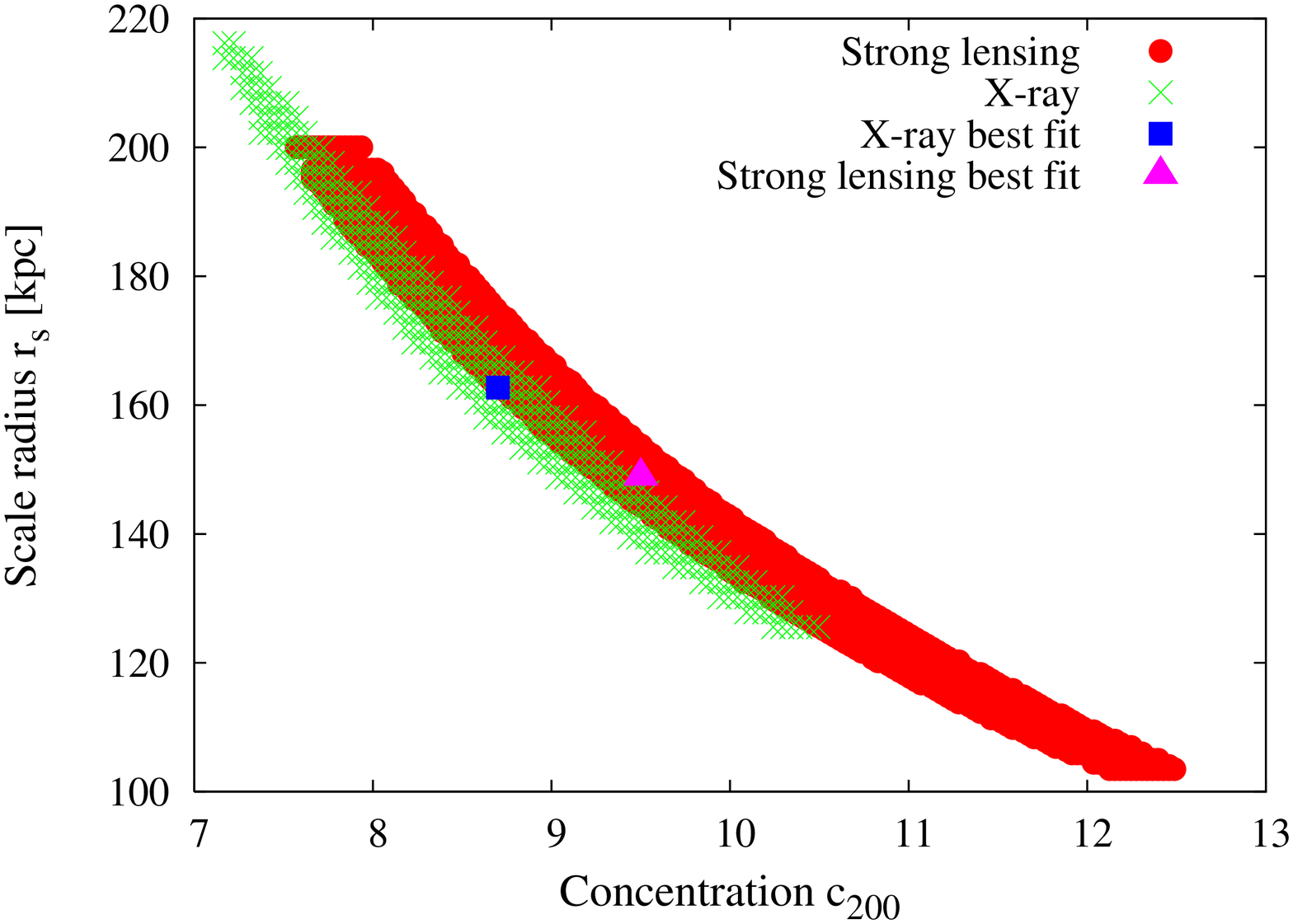,width=0.45\textwidth,angle=-90} & \psfig{figure=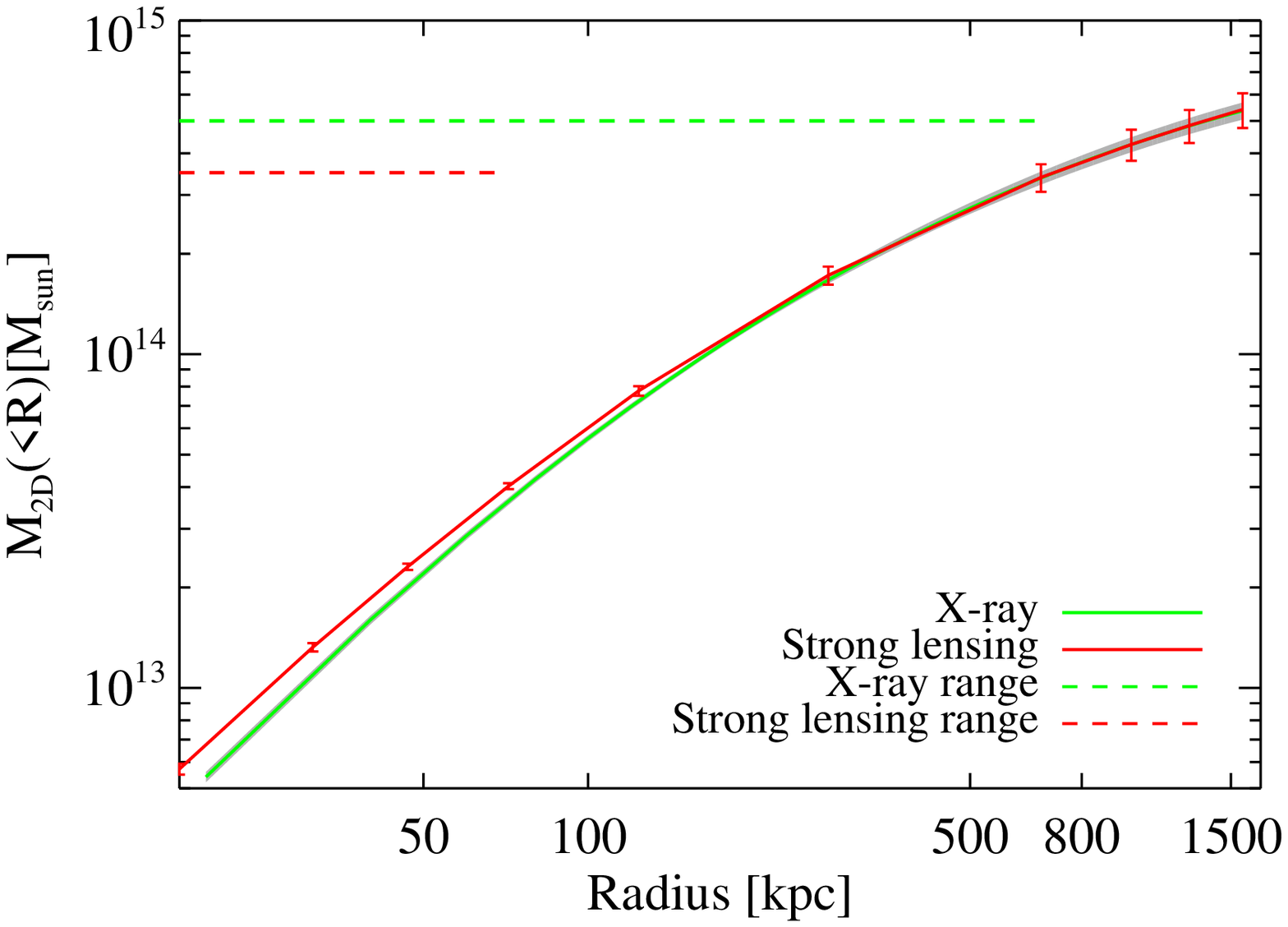,width=0.45\textwidth}
\end{tabular}
 \caption[]{[\emph{Left panel}] $3\sigma$  marginalized probability distribution for the NFW parameters - concentration and scale radius. The regions  refer to the the X-ray (drawn in green) and strong lensing (drawn in red) results. The square [triangle] indicates the X-ray [strong lensing] best fit values. [\emph{Right panel}] Projected total mass enclosed in cylinders of radius R. The solid green line represents the  projected total mass profile derived through the X-ray analysis (the grey area indicates the $1\sigma$ region), the red line with error bars show the result of the strong lensing analysis. The error bars indicate the $1\sigma$ errors inferred from the Bayesian optmization. The green [red] dashed horizontal line indicates the spatial range over which the X-ray [strong lensing] analysis has been performed.}
 \label{fig:cont_both}  
 \end{center} 
\end{figure*}

Coupling the X-ray and lensing mass estimates is not a trivial task, as already stated by G05, due to their different dependencies on the characteristics of galaxy clusters (like asphericity) as well as for the still incomplete treatment of the biases that could affect both measurements. Though joint analyses of homogeneous effects (e.g. weak and strong lensing) often supply converging estimates, leading to remarkable results \citep{bradac2005,bradac2006,limousin2007,merten2008},  comparing 2- and  3-dimensional measurements  requires further caution. Recently, several interesting attempts have been made to infer cluster masses through multi-wavelength, combined analysis \citep{mahdavi2007}, in some cases suggesting that a direct comparison needs further investigation \citep{lemze2008}. We decided instead to perform two parallel analyses to understand if these methods lead to discrepant mass estimates for MS2137. \\
%----------
\begin{table*}  
\begin{center} 
\caption[]{Best fit  parameters for the NFW profile. The strong lensing results refer to the reference model, labeled [1] in Tab. \ref{tab:lens_comp}. Quoted errors correspond to $1\sigma$ confidence level. For a comparison, we reported the results obtained in some earlier representative studies:  the original values  were converted according to common definitions. We list here the findings obtained by \cite{allen2001}  and \cite{schmidt2007} through the X-ray  analysis of \textit{Chandra} data (columns A01 and S07, respectively) and  by \cite{gavazzi2005} and \cite{comerford2006}  through a strong lensing study (columns G05 and C06, respectively). Here, again, $e=(a^{2}-b^{2})/(a^{2}+b^{2})$.}
\label{tab:tab_res}  
\input{tab_res1.tex}

\end{center}
\end{table*}  
%----------
We found that our X-ray and strong lensing results are  in good agreement: both the extrapolated value of the total mass $M_{200}$ and the mass model parameters agree within the $2\sigma$ range.  The parameter  probability distributions are mutually consistent (see the left panel of Figure \ref{fig:cont_both}), although the concentration inferred from the lensing analysis is slightly higher than the X-ray one. A mild elongation of the cluster along the line of sight could be a possible explanation, as already suggested by G05. Another reason could be the additional uncertainty associated to the BCG mass. To evaluate the differences between the X-ray and the strong lensing results, we show in the right panel of Figure \ref{fig:cont_both} the projected total mass enclosed in concentric cylinders inferred from both analyses. The assessment of the errors on the strong lensing mass was performed by stacking 500 mass map realizations, obtained through the $\emph{Lenstool}$ Bayesian optimization, and computing the standard deviation at each map point.

\section{Comparison with previous studies}
\label{sec:prev_comp}
In this section we will briefly compare our results with some  previous studies, namely the findings obtained  by \citet*{schmidt2007} (hereafter S07) and \citet*{allen2001} (hereafter A01) through the X-ray analysis of \textit{Chandra} data, and those inferred through a strong lensing analysis by  G05 and  \citet{comerford2006} (hereafter C06: see also \citealp{gavazzi2003,comerford2007,shu2008}). \\
S07 studied the same two \textit{Chandra} datasets considered in this work, whereas the analysis of A01 was carried out on the only \textit{Chandra} dataset  available at that epoch (Obs. 928). The method applied in their studies are similar to ours. Briefly, S07 and A01  derived the observed X-ray surface brightness profile and the deprojected X-ray gas temperature profile, and combined them to determine the gas mass and total mass profile of the galaxy cluster, under the assumptions of hydrostatic equilibrium and spherical symmetry\footnote{S07 carried out two separate mass analyses, deriving  a mass estimate of the cluster total matter and of its three components (dark matter, gas and cD galaxy). In the following, we will refer to the results  obtained in the first case.} (for more details, see also \citealt*{white1997,allen2001b}). The gas temperature and density profiles for the single case of MS2137 were not listed in S07 and A01,  since in both cases the authors  presented the results of careful analyses of large samples of objects. Thus a direct comparison of the derived profiles was not possible. 
Concerning the final estimates of the NFW parameters,  there is a good matching between our results and the values presented both in S07 and A01. Despite some differences in the analyses (for example the spectral binning and the X-ray background estimate) the results are mutually  consistent within a $2\sigma$ level (see  Tab.\ref{tab:tab_res}). \\
 A comparison with the previous strong lensing analyses involves an extended discussion, since, as already stated by \cite{comerford2007}, the scatter in the differing results found in the literature regarding this cluster is quite large.  A complete discussion on this topic is beyond the aim of this paper: here we will  simply compare the results and  the methods of analysis, and possibly suggest some explanations. For full details on the basis of the optimization methods,  the reader can refer  to the related papers,   \citet{gavazzi2005} and  \citet{comerford2006} respectively. \\
Both  the analyses of G05 and C06 were performed  exploiting their own  inversion codes.
 The source plane $\chi^2$ estimator adopted in  G05 follows the $\chi^2_{\rm s}$ definition in  the \texttt{lensmodel} software \citep{keeton2001b,keeton2001a}.
 G05  reported a value for the scale radius $r_{s}$  consistent with our estimate, whereas there is  disagreement between our estimates of the halo concentration and mass. We couldn't find a convincing explanation for the difference in our results. \\
The  code utilised in C06 optimizes the lens model parameters  by minimizing  a figure-of-merit function, defined as the sum of three components both in the image and in the source planes. The  optimization procedure is meant to shrink:
\begin{inparaenum}
\item  the distance between each data point position  and the nearest predicted one,
\item the distance between the  predicted point positions and the nearest data point (thus penalising models that produce unobserved images) and
\item the size and  noncompactness of the predicted sources (see C06 and \citealp{shu2008} for a detailed discussion).
\end{inparaenum}

This fitting technique is very accurate: however, in their analysis C06 modeled the lens as a single elliptical NFW potential, not  including any galaxy component. Thus, the free parameters of their lens model are the scale radius $\rm r_{\rm s}$, the ellipticity, the position angle $\theta$ and the scale convergence $\rm k_{\rm s}\equiv \rho_{\rm s}r_{\rm s}/\Sigma_\mathrm{crit}$, where as usual $\Sigma_\mathrm{crit}$ represents the critical surface mass density.
The resulting MS2137 mass parameters   are $\rm r_{\rm s} \simeq 91.4\pm3$ kpc and  $\rm c\simeq13.0\pm1.0$, leading to  a total mass extrapolated up to $\rm R_{200}$ of  $\rm M_{200}\simeq 2.9\pm0.4\times 10^{14}\rm \ M_{\odot}$. Thus, the total mass  and  scale radius estimates of C06 are lower and their inferred concentration parameter is higher than our best-fit values. \\
 We performed a single mass component analysis, in analogy with  C06 and \cite{shu2008}.  The best-fit values we obtained in this case are:  
\begin{enumerate}
\item  $\rm r_{s} = 92.8\pm 10.0$ kpc,
\item  $\rm c = 13.5 \pm 1.1$,
\item  $\rm M_{200} = 3.1 \pm 0.3\times 10^{14}\rm \ M_{\odot}$,
\end{enumerate}
which are in excellent agreement with the C06 values; the total $\chi^2$  is $29.6$ for 63 d.o.f.  This result highlights  once again that the BCG  has a non-negligible effect on strong lensing properties \citep{hilbert2008} when dealing with lensed features  sensitive to the  central mass distribution, such as radial arcs.  Ignoring the BCG stellar mass component can return an ``artificial'' high-concentrated profile, to account for the missing central galactic component.

\section{Discussions and conclusions}
\label{sec:concl}
We have presented new X-ray and strong lensing mass measurements for the galaxy cluster MS2137.3-2353, taking advantages of the high-resolution \textit{Chandra} and \emph{HST} data. Our X-ray analysis benefits from a non-parametric study of the gas temperature and density profiles. By combining them, we have recovered the cluster mass profile adopting the NFW model, under the assumptions of  spherical symmetry and hydrostatic equilibrium. \\
 The strong lensing  reconstruction was performed through the publicly available code \emph{Lenstool}   \citep{kneib1996,jullo2007}, which allowed for a parametric analysis comparable to our X-ray study. 
 Our mass estimates are consistent within the errors, leading to a mean value of the total mass, extrapolated up to $\rm R_{200}$, of $\simeq 4.4 \pm 0.3\times 10^{14}\rm M_{\odot}$. We did not find a strong discrepancy between the strong lensing and the X-ray NFW parameter results, since the best-fit values mutually agree at a $3\sigma$ confidence level. The probability distributions are mutually compatible.  The slight difference in the allowed parameter space could be explained,  for example, by a mild cluster elongation along the line of sight, which would affect mainly the strong lensing results. Moreover, the uncertainty on the BCG mass determination could introduce a systematic bias in the lensing measurements that we could quantify as an additional  uncertainty of $\simeq 25\%$, since the extrapolated strong lensing mass estimate in this case depends upon the BCG stellar mass budget. \\
The agreement between these two mass estimates for a relaxed cluster, like MS2137, supports the statement that, for this kind of  system, both the strong lensing and the X-ray analyses provide an unbiased way to estimate the cluster mass. As a general observation on the physics of this cluster, the convergence of these mass results implies that its main pressure support mechanism against gravity is the thermal pressure, so the assumption of hydrostatic equilibrium is verified. Non-thermal processes, such as turbulent motions or magnetic field effects, seemingly have little impact on the equilibrium state for this cluster, that appears a good example of relaxed cluster prototype. To verify if MS2137 represents a peculiar case in the cluster   mass discrepancy debate, or if the relaxed dynamical state constitutes a sufficient  condition for the convergence of the strong lensing and X-ray mass estimates, further investigations of both single cluster cases and large uniform samples are  fundamental, and would provide new, useful insights on the galaxy cluster physics. 

\section{acknowledgments}
\label{sec:ackno}
 We would like to thank  Rapha\"el Gavazzi, Bernard Fort, Mario Radovich, Fabio Gastaldello, Giovanni Covone, Eric Jullo and Marceau Limousin for their help and suggestions. We are grateful to Matthias Bartelmann for reading the paper and providing useful comments. We thank the anonymous referee for useful comments and remarks that improved the presentation of our work. AD  acknowledges the support of European Association for Research in Astronomy (MEST-CT-2004-504604 Marie Curie - EARA EST fellowship), and  thanks Rapha\"el Gavazzi and Bernard Fort for their kind hospitality during the visit at  IAP - Institut d'Astrophysique de Paris. AD also thanks Daniela Crociani  and Ashley J. Ruiter for their help and support. We thank Jean Paul Kneib and the \emph{Lenstool} developers for making their lensing software public. This research has made use of data obtained from the \textit{Chandra} Data Archive and of the software provided by the \textit{Chandra} X-ray Center (CXC), and of \emph{HST} data obtained from the MAST archive. We acknowledge the financial contribution from contracts ASI-INAF I/023/05/0, I/088/06/0 and I/016/07/0.
\bibliography{master2}
\end{document}

%% file: specres.tex
\begin{tabular}{@{\@{\hspace{1pt}}}lc@{\@{\hspace{3pt}}}c@{\@{\hspace{4pt}}}c@{\@{\hspace{4pt}}}c@{\@{\hspace{4pt}}}c@{\@{\hspace{4pt}}}c}
\multicolumn{5}{c}{\textsc{Spectral fit results}} \\
\hline \hline
$\rm R_{out}$ & $\rm T_{\rm gas}$  & Normalization & Metal abundance &  Goodness of fit \\
$\rm [arcmin] $& $\rm [keV]$ & $\times 10^{4}$& $[Z_{\odot}]$&  \\ 
\hline
$0.065$  &$ 4.33_{-0.19}^{+0.20}$  &$  7.03_{-0.27}^{+0.25}$ &     $0.60_{-0.11}^{+0.13}$ & $  459.2 [567]$ \\
$ 0.114$  & $4.22_{-0.18}^{+0.19}$ &   $7.62_{-0.29}^{+0.28}$ & $0.58_{-0.11}^{+0.13} $  & $514.8 [552]$ \\
  $0.17$ &  $ 4.66_{-0.21}^{+0.23}$ & $  7.43_{-0.26}^{+0.25}$ &   $0.52_{-0.11}^{+0.12}$ & $578.7 [579]$ \\
$0.26$ &  $ 5.35_{-0.26}^{+0.28} $& $ 8.67_{-0.27}^{+0.25} $& $0.50_{-0.10}^{+0.12}$ &  $597.0 [615]$\\
$ 0.45$ & $  5.88_{-0.28}^{+0.29} $&$   11.17_{-0.30}^{+0.28}$  &$ 0.41_{-0.09}^{+0.10}$&$ 670.3 [679]$\\
 $1.0$ &$   5.03_{-0.23}^{+0.25} $&$   12.15_{-0.33}^{+0.32}$ & $ 0.38_{-0.09}^{+0.10}$ & $658.3 [740]$\\ 
$ 2.45^{a}$ & $  3.63_ {-0.43}^{+0.59} $ & $   6.50_{-0.56}^{+0.54}$ & $ 0.40_{-0.21}^{+0.30}$   & $  348.0 [291]$ \\
\hline
\end{tabular}

%% file: tab_knotsa.tex
\begin{tabular*}{0.48\textwidth}{l|r@{\@{\hspace{6pt}}}r@{\@{\hspace{6pt}}}r@{\@{\hspace{6pt}}}r@{\@{\hspace{6pt}}}r@{\@{\hspace{6pt}}}r@{\@{\hspace{6pt}}}r@{\@{\hspace{6pt}}}r}
\multicolumn{9}{c}{\textsc{Multiple-image systems}} \\
\hline 
Id & \multicolumn{2}{c}{Im.1} & \multicolumn{2}{c}{Im.2} &  \multicolumn{2}{c}{Im.3} &  \multicolumn{2}{c}{Im.4   } \\
\hline \hline
1.a & -5.20 & 13.87 & 2.61 & 14.86 & 13.59 & -1.08 & -12.03 & -15.29 \\
1.b & -4.88 & 13.98 & 2.31 & 14.89 &  13.46 & -1.26 & -12.36 & -14.94   \\
1.c &  -7.32 & 13.68 & 0.04 & 15.77  & 13.06 & -2.96  & -13.32 & -13.63 \\
1.d & -7.45 & 12.74 & 4.69 & 14.80  & 13.62 & -2.13 & -11.83 & -15.00 \\
1.e & -3.79 & 14.52  & 0.27 & 15.06  & 13.33 & -1.40 & -12.62 & -14.71 \\
1.f & -6.85 & 13.86  & -1.35 & 15.71  &  12.93 & -2.78 & -13.51 & -13.68   \\
1.g & -7.81 & 13.26  &  1.32 & 15.82 & 12.86 & -3.28   & -13.32 & -13.50 \\ % 1.a
1.h & -5.70 & 14.55   & -2.14 & 15.60 &  12.80 & -3.06 & -13.55 & -13.53   \\ % 1.b
1.i & & & & & 13.13 & -1.23 & -13.28 & -14.45   \\ %1.c
1.j & & & & & 12.01 & -2.12 & -15.13 & -12.80 \\ %1.8
1.k & & & & & -14.17 & -13.40  & 12.50 & -2.01   \\ %1.9
\hline
2.a & 0.04 & 6.79 & 0.33 & 3.41  & -7.28 & -22.49 & & \\
2.b & 0.00 & 6.44  & 0.27 & 3.88  & -7.22 & -22.75 & & \\
2.c & -0.06 & 5.87 & 0.17 & 4.21  & -7.05 & -22.88 & & \\
2.d & -0.03 & 5.28 & 0.10 & 4.50  & -6.92 & -23.04 & & \\
\hline
\end{tabular*}

%% file: tab_comp.tex
\begin{tabular}{@{\@{\hspace{1pt}}}l|c@{\@{\hspace{1pt}}}c@{\@{\hspace{3pt}}}c@{\@{\hspace{4pt}}}c@{\@{\hspace{4pt}}}c@{\@{\hspace{4pt}}}c@{\@{\hspace{4pt}}}c@{\@{\hspace{4pt}}}c@{\@{\hspace{4pt}}}c@{\@{\hspace{1pt}}}}
 \multicolumn{10}{c}{\textsc{Best fit strong lensing results}} \\
\hline 
 Model & $\rm x_{ c} $ & $\rm y_{c}$ & $ e$ & $\theta$  & $\rm r_{\rm s}$& $\rm c_{200}$ & $\rm M_{200}$ & $\sigma_{0}$ & $\chi^{2}_{\rm tot} [\rm d.o.f.]$\\
  & [arcsec] &  [arcsec] & & [deg] & [kpc] & & $[\rm 10^{14} M_{\odot}]$ & [km $\rm s^{-1}$] &  \\
\hline
\hline
$[1]$ \small{NFW+dPIE} & $0.0$ & $0.0$  &  $0.10$ & $147.2^{+0.6}_{-0.4} [147.4]$  & $148.9^{+22.0}_{-24.7} [143.5]$ & $9.50^{+1.29}_{-0.87} [9.76]$ &$ 4.11^{+0.75}_{-0.46} [4.31]$ & $1377.3^{+26.9}_{-36.2} [1366.4]$ & $17.6 [61]$\\
\hline
$[2]$ \small{NFW+SIE} & $0.1$ & $-0.1$ & $0.10$ & $149.3\pm0.7 [149.2]$ & $118.3^{+26.0}_{-14.3} [117.3]$ & $10.47^{+1.26}_{-1.23} [10.98]$ & $3.38^{+0.59}_{-0.35} [3.36]$ & $1291.2^{+28.9}_{-20.5}$ &  $10.9 [61]$ \\
\hline
\end{tabular}

%% file: best_pot.tex
\begin{tabular}{lcccc}
\multicolumn{5}{c}{\textsc{Galaxy model parameters}} \\
\hline
Model  & Halo &	Ellipticity    & $\rm R_{\rm cut}$ & $\sigma_{0}$ \\
            &         &                         & $[\rm kpc]$            &  $ [\rm km\ s^{-1}]$ \\
 \hline
 \hline
$[1]   $      &  BCG     &  $[0.18]$   &   $[22.0]$  &  $230.7^{+11.2}_{-0.49} (230.5)$ \\
             & Galaxy  & 	 $[0.38]$ & $ [5.0]$  & $150.2^{+26.5}_{-23.9} (159.2)$ \\
$[2]$         &  BCG     & $ [0.18] $  & - &  $ 231.4^{+33.3}_{-1.0} (240.3)$ \\
             & Galaxy  & 	 $[0.38] $ & - & $124.5^{+12.9}_{-16.5} (124.2)$ \\
\hline
\end{tabular}   

%% file: tab_chi2.tex
\begin{tabular}{l|ccc}
 \multicolumn{4}{c}{\textsc{Goodness of fit}} \\
\hline 
System & $\chi^2$ & $\rm rms_s ['']$ & $\rm rms_i ['']$ \\
\hline \hline
1.a   & 0.48    &  0.034  & 0.27   \\
1.b   &  1.72   &  0.064  & 0.37  \\
1.c   &  2.18   &  0.075  & 0.22    \\
1.d   &  0.90   &  0.053  & 0.23   \\
1.e   &  1.80   &  0.062  & 0.50   \\
1.f   &  1.61   &  0.062  & 0.86 \\
1.g   &  1.06   &  0.057  & 0.37   \\
1.h   &  0.21   &  0.021  & 0.25  \\
1.i   &  0.11   &  0.030  & 0.08   \\
1.l   &  0.16   &  0.038  & 0.10  \\
1.m   &  0.14   &  0.034  & 0.07   \\
\hline
2.a   &  3.40   &  0.168 & 0.33  \\
2.b   &  1.97   &  0.114 & 0.35  \\
2.c   &  1.13   &  0.078 & 0.28  \\
2.d   &  0.75   &  0.057 & 0.44    \\
\hline
Total & $\chi^2$ [d.o.f] & $\rm rms_s ['']$ & $\rm rms_i ['']$ \\
      &  17.62 [61] &  0.072 & 0.39  \\
\hline    
\end{tabular}

%% file: tab_res1.tex
\begin{tabular}{l|ccccccccc}
\multicolumn{8}{c}{\textsc{Comparison with previous results}} \\
\hline 
 & X-RAY  & SL   & A01 & S07 & G05 & C06  & Units \\
\hline
\hline
 $\rm r_{s}$  &  $ 162.72^{+18.60}_{-16.27}$  & $148.9^{+22.0}_{-24.7}$ & $160.0\pm30$ &   $180.0\pm20.0$ &$158.0^{+15}_{-13}$ &  $91.4\pm3$  &  $\rm kpc$ \\
 $\rm c_{200}$ &    $   8.68^{+0.71}_{-0.91}$ & $9.50^{+1.29}_{-0.87}$ & $8.71_{-0.92}^{+1.22}$&  $8.19^{+0.54}_{-0.56}$ &$11.92^{+0.77}_{-0.74}$ & $13.0\pm1.0$  & - \\
  %$\rm r_{200}$ & $ 1.37\pm0.04$ & $1.41$ & $1.39$ & $1.88 \pm0.05$ & $0.83$ &  $\rm Mpc$ \\
 $\rm M_{200}$   &   $4.43\pm{0.25}$  & $4.11^{+0.75}_{-0.46}$ &  - & - &$7.56^{+0.63}_{-0.54}$  & $2.9\pm0.4$  & $10^{14} \rm M_{\odot}$ \\
 $\emph{e}$ & - & $0.1\pm{0.01} $ & - &-& $0.25$ & $0.12$& - \\
\hline
\end{tabular}

%% file: ms.bbl
\newcommand{\noopsort}[1]{}
\begin{thebibliography}{}

\bibitem[\protect\citeauthoryear{{Allen}}{{Allen}}{1998}]{allen1998}
{Allen} S.~W.,  1998, \mnras, 296, 392

\bibitem[\protect\citeauthoryear{{Allen}, {Ettori} \& {Fabian}}{{Allen}
  et~al.}{2001}]{allen2001b}
{Allen} S.~W.,  {Ettori} S.,    {Fabian} A.~C.,  2001, \mnras, 324, 877

\bibitem[\protect\citeauthoryear{{Allen}, {Schmidt} \& {Fabian}}{{Allen}
  et~al.}{2001}]{allen2001}
{Allen} S.~W.,  {Schmidt} R.~W.,    {Fabian} A.~C.,  2001, \mnras, 328, L37

\bibitem[\protect\citeauthoryear{{Arnaud}}{{Arnaud}}{1996}]{arnaud1996}
{Arnaud} K.~A.,  1996, in {Jacoby} G.~H.,  {Barnes} J.,  eds, Astronomical Data
  Analysis Software and Systems V Vol.~101 of Astronomical Society of the
  Pacific Conference Series, {XSPEC: The First Ten Years}.
pp 17--+

\bibitem[\protect\citeauthoryear{{Arnaud}, {Pointecouteau} \& {Pratt}}{{Arnaud}
  et~al.}{2005}]{arnaud2005}
{Arnaud} M.,  {Pointecouteau} E.,    {Pratt} G.~W.,  2005, \aap, 441, 893

\bibitem[\protect\citeauthoryear{{Bartelmann}}{{Bartelmann}}{1996}]{bartelmann%
1996}
{Bartelmann} M.,  1996, \aap, 313, 697

\bibitem[\protect\citeauthoryear{{Bartelmann} \& {Narayan}}{{Bartelmann} \&
  {Narayan}}{1995}]{bartelmann1995}
{Bartelmann} M.,  {Narayan} R.,  1995, in {Holt} S.~S.,  {Bennett} C.~L.,  eds,
  Dark Matter Vol.~336 of American Institute of Physics Conference Series,
  {Gravitational Lensing and the Mass Distribution of Clusters}.
pp 307--+

\bibitem[\protect\citeauthoryear{{Borgani}}{{Borgani}}{2006}]{borgani2006}
{Borgani} S.,  2006, ArXiv e-prints, astro-ph/0605575

\bibitem[\protect\citeauthoryear{{Brada{\v c}}, {Clowe}, {Gonzalez},
  {Marshall}, {Forman}, {Jones}, {Markevitch}, {Randall}, {Schrabback} \&
  {Zaritsky}}{{Brada{\v c}} et~al.}{2006}]{bradac2006}
{Brada{\v c}} M.,  {Clowe} D.,  {Gonzalez} A.~H.,  {Marshall} P.,  {Forman} W.,
   {Jones} C.,  {Markevitch} M.,  {Randall} S.,  {Schrabback} T.,    {Zaritsky}
  D.,  2006, \apj, 652, 937

\bibitem[\protect\citeauthoryear{{Brada{\v c}}, {Erben}, {Schneider},
  {Hildebrandt}, {Lombardi}, {Schirmer}, {Miralles}, {Clowe} \&
  {Schindler}}{{Brada{\v c}} et~al.}{2005}]{bradac2005}
{Brada{\v c}} M.,  {Erben} T.,  {Schneider} P.,  {Hildebrandt} H.,  {Lombardi}
  M.,  {Schirmer} M.,  {Miralles} J.-M.,  {Clowe} D.,    {Schindler} S.,  2005,
  \aap, 437, 49

\bibitem[\protect\citeauthoryear{{Brada{\v c}}, {Schrabback}, {Erben},
  {McCourt}, {Million}, {Mantz}, {Allen}, {Blandford}, {Halkola},
  {Hildebrandt}, {Lombardi}, {Marshall}, {Schneider}, {Treu} \&
  {Kneib}}{{Brada{\v c}} et~al.}{2008}]{bradac2008}
{Brada{\v c}} M.,  {Schrabback} T.,  {Erben} T.,  {McCourt} M.,  {Million} E.,
  {Mantz} A.,  {Allen} S.,  {Blandford} R.,  {Halkola} A.,  {Hildebrandt} H.,
  {Lombardi} M.,  {Marshall} P.,  {Schneider} P.,  {Treu} T.,    {Kneib} J.-P.,
   2008, \apj, 681, 187

\bibitem[\protect\citeauthoryear{{Buote}}{{Buote}}{2000}]{buote2000}
{Buote} D.~A.,  2000, \apj, 539, 172

\bibitem[\protect\citeauthoryear{{Comerford}, {Meneghetti}, {Bartelmann} \&
  {Schirmer}}{{Comerford} et~al.}{2006}]{comerford2006}
{Comerford} J.~M.,  {Meneghetti} M.,  {Bartelmann} M.,    {Schirmer} M.,  2006,
  \apj, 642, 39

\bibitem[\protect\citeauthoryear{{Comerford} \& {Natarajan}}{{Comerford} \&
  {Natarajan}}{2007}]{comerford2007}
{Comerford} J.~M.,  {Natarajan} P.,  2007, \mnras, 379, 190

\bibitem[\protect\citeauthoryear{{Dalal} \& {Keeton}}{{Dalal} \&
  {Keeton}}{2003}]{dalal2003}
{Dalal} N.,  {Keeton} C.~R.,  2003, ArXiv e-prints, astro-ph/0312072

\bibitem[\protect\citeauthoryear{{Diaferio}, {Schindler} \& {Dolag}}{{Diaferio}
  et~al.}{2008}]{diaferio2008}
{Diaferio} A.,  {Schindler} S.,    {Dolag} K.,  2008, Space Science Reviews,
  134, 7

\bibitem[\protect\citeauthoryear{{Dickey} \& {Lockman}}{{Dickey} \&
  {Lockman}}{1990}]{dickey1990}
{Dickey} J.~M.,  {Lockman} F.~J.,  1990, \araa, 28, 215

\bibitem[\protect\citeauthoryear{{El{\'{\i}}asd{\'o}ttir}, {Limousin},
  {Richard}, {Hjorth}, {Kneib}, {Natarajan}, {Pedersen}, {Jullo} \&
  {Paraficz}}{{El{\'{\i}}asd{\'o}ttir} et~al.}{2007}]{eliasdottir2007}
{El{\'{\i}}asd{\'o}ttir} {\'A}.,  {Limousin} M.,  {Richard} J.,  {Hjorth} J.,
  {Kneib} J.-P.,  {Natarajan} P.,  {Pedersen} K.,  {Jullo} E.,    {Paraficz}
  D.,  2007, ArXiv e-prints, astro-ph/07105636, 710

\bibitem[\protect\citeauthoryear{{Ettori}, {De Grandi} \& {Molendi}}{{Ettori}
  et~al.}{2002}]{ettori2002}
{Ettori} S.,  {De Grandi} S.,    {Molendi} S.,  2002, \aap, 391, 841

\bibitem[\protect\citeauthoryear{{Ettori} \& {Lombardi}}{{Ettori} \&
  {Lombardi}}{2003}]{ettori2003}
{Ettori} S.,  {Lombardi} M.,  2003, \aap, 398, L5

\bibitem[\protect\citeauthoryear{{Fort}, {Le Fevre}, {Hammer} \&
  {Cailloux}}{{Fort} et~al.}{1992}]{fort1992}
{Fort} B.,  {Le Fevre} O.,  {Hammer} F.,    {Cailloux} M.,  1992, \apjl, 399,
  L125

\bibitem[\protect\citeauthoryear{{Gao}, {Navarro}, {Cole}, {Frenk}, {White},
  {Springel}, {Jenkins} \& {Neto}}{{Gao} et~al.}{2008}]{gao2008}
{Gao} L.,  {Navarro} J.~F.,  {Cole} S.,  {Frenk} C.~S.,  {White} S.~D.~M.,
  {Springel} V.,  {Jenkins} A.,    {Neto} A.~F.,  2008, \mnras, 387, 536

\bibitem[\protect\citeauthoryear{{Gavazzi}}{{Gavazzi}}{2005a}]{gavazzi2005a}
{Gavazzi} R.,  2005a, in {Mellier} Y.,  {Meylan} G.,  eds, Gravitational
  Lensing Impact on Cosmology Vol.~225 of IAU Symposium, {Lensing \& Dynamics
  in the Galaxy Cluster MS2137-23}.
pp 179--184

\bibitem[\protect\citeauthoryear{{Gavazzi}}{{Gavazzi}}{2005b}]{gavazzi2005}
{Gavazzi} R.,  2005b, \aap, 443, 793

\bibitem[\protect\citeauthoryear{{Gavazzi}, {Fort}, {Mellier}, {Pell{\'o}} \&
  {Dantel-Fort}}{{Gavazzi} et~al.}{2003}]{gavazzi2003}
{Gavazzi} R.,  {Fort} B.,  {Mellier} Y.,  {Pell{\'o}} R.,    {Dantel-Fort} M.,
  2003, \aap, 403, 11

\bibitem[\protect\citeauthoryear{{Gitti}, {Piffaretti} \& {Schindler}}{{Gitti}
  et~al.}{2007}]{gitti2007}
{Gitti} M.,  {Piffaretti} R.,    {Schindler} S.,  2007, \aap, 472, 383

\bibitem[\protect\citeauthoryear{{Golse} \& {Kneib}}{{Golse} \&
  {Kneib}}{2002}]{golse2002}
{Golse} G.,  {Kneib} J.-P.,  2002, \aap, 390, 821

\bibitem[\protect\citeauthoryear{{Halkola}, {Hildebrandt}, {Schrabback},
  {Lombardi}, {Brada{\v c}}, {Erben}, {Schneider} \& {Wuttke}}{{Halkola}
  et~al.}{2008}]{halkola2008}
{Halkola} A.,  {Hildebrandt} H.,  {Schrabback} T.,  {Lombardi} M.,  {Brada{\v
  c}} M.,  {Erben} T.,  {Schneider} P.,    {Wuttke} D.,  2008, \aap, 481, 65

\bibitem[\protect\citeauthoryear{{Hallman}, {Motl}, {Burns} \&
  {Norman}}{{Hallman} et~al.}{2006}]{hallman2006}
{Hallman} E.~J.,  {Motl} P.~M.,  {Burns} J.~O.,    {Norman} M.~L.,  2006, \apj,
  648, 852

\bibitem[\protect\citeauthoryear{{Hilbert}, {White}, {Hartlap} \&
  {Schneider}}{{Hilbert} et~al.}{2008}]{hilbert2008}
{Hilbert} S.,  {White} S.~D.~M.,  {Hartlap} J.,    {Schneider} P.,  2008,
  \mnras, pp 471--+

\bibitem[\protect\citeauthoryear{{Jing} \& {Suto}}{{Jing} \&
  {Suto}}{2000}]{jing2000}
{Jing} Y.~P.,  {Suto} Y.,  2000, \apjl, 529, L69

\bibitem[\protect\citeauthoryear{{Jullo}, {Kneib}, {Limousin},
  {El{\'{\i}}asd{\'o}ttir}, {Marshall} \& {Verdugo}}{{Jullo}
  et~al.}{2007}]{jullo2007}
{Jullo} E.,  {Kneib} J.-P.,  {Limousin} M.,  {El{\'{\i}}asd{\'o}ttir} {\'A}.,
  {Marshall} P.~J.,    {Verdugo} T.,  2007, New Journal of Physics, 9, 447

\bibitem[\protect\citeauthoryear{{Kaastra} \& {Mewe}}{{Kaastra} \&
  {Mewe}}{1993}]{kaastra1993}
{Kaastra} J.~S.,  {Mewe} R.,  1993, \aaps, 97, 443

\bibitem[\protect\citeauthoryear{{Keeton}}{{Keeton}}{2001a}]{keeton2001b}
{Keeton} C.~R.,  2001a, ArXiv e-prints, astro-ph/0102341

\bibitem[\protect\citeauthoryear{{Keeton}}{{Keeton}}{2001b}]{keeton2001a}
{Keeton} C.~R.,  2001b, ArXiv e-prints, astro-ph/0102340

\bibitem[\protect\citeauthoryear{{Keeton}, {Kochanek} \& {Seljak}}{{Keeton}
  et~al.}{1997}]{keeton1997}
{Keeton} C.~R.,  {Kochanek} C.~S.,    {Seljak} U.,  1997, \apj, 482, 604

\bibitem[\protect\citeauthoryear{{King}}{{King}}{1962}]{king1962}
{King} I.,  1962, \aj, 67, 274

\bibitem[\protect\citeauthoryear{{King}}{{King}}{1966}]{king1966}
{King} I.~R.,  1966, \aj, 71, 64

\bibitem[\protect\citeauthoryear{{Kneib}, {Ellis}, {Smail}, {Couch} \&
  {Sharples}}{{Kneib} et~al.}{1996}]{kneib1996}
{Kneib} J.-P.,  {Ellis} R.~S.,  {Smail} I.,  {Couch} W.~J.,    {Sharples}
  R.~M.,  1996, \apj, 471, 643

\bibitem[\protect\citeauthoryear{{Kriss}, {Cioffi} \& {Canizares}}{{Kriss}
  et~al.}{1983}]{kriss1983}
{Kriss} G.~A.,  {Cioffi} D.~F.,    {Canizares} C.~R.,  1983, \apj, 272, 439

\bibitem[\protect\citeauthoryear{{Lemze}, {Barkana}, {Broadhurst} \&
  {Rephaeli}}{{Lemze} et~al.}{2008}]{lemze2008}
{Lemze} D.,  {Barkana} R.,  {Broadhurst} T.~J.,    {Rephaeli} Y.,  2008,
  \mnras, 386, 1092

\bibitem[\protect\citeauthoryear{{Limousin}, {Kneib} \& {Natarajan}}{{Limousin}
  et~al.}{2005}]{limousin2005}
{Limousin} M.,  {Kneib} J.-P.,    {Natarajan} P.,  2005, \mnras, 356, 309

\bibitem[\protect\citeauthoryear{{Limousin}, {Richard}, {Jullo}, {Kneib},
  {Fort}, {Soucail}, {El{\'{\i}}asd{\'o}ttir}, {Natarajan}, {Ellis}, {Smail},
  {Czoske}, {Smith}, {Hudelot}, {Bardeau}, {Ebeling}, {Egami} \&
  {Knudsen}}{{Limousin} et~al.}{2007}]{limousin2007}
{Limousin} M.,  {Richard} J.,  {Jullo} E.,  {Kneib} J.-P.,  {Fort} B.,
  {Soucail} G.,  {El{\'{\i}}asd{\'o}ttir} {\'A}.,  {Natarajan} P.,  {Ellis}
  R.~S.,  {Smail} I.,  {Czoske} O.,  {Smith} G.~P.,  {Hudelot} P.,  {Bardeau}
  S.,  {Ebeling} H.,  {Egami} E.,    {Knudsen} K.~K.,  2007, \apj, 668, 643

\bibitem[\protect\citeauthoryear{{Mahdavi}, {Hoekstra}, {Babul}, {Sievers},
  {Myers} \& {Henry}}{{Mahdavi} et~al.}{2007}]{mahdavi2007}
{Mahdavi} A.,  {Hoekstra} H.,  {Babul} A.,  {Sievers} J.,  {Myers} S.~T.,
  {Henry} J.~P.,  2007, \apj, 664, 162

\bibitem[\protect\citeauthoryear{{Markevitch}, {Bautz}, {Biller}, {Butt},
  {Edgar}, {Gaetz}, {Garmire}, {Grant}, {Green}, {Juda}, {Plucinsky},
  {Schwartz}, {Smith}, {Vikhlinin}, {Virani}, {Wargelin} \&
  {Wolk}}{{Markevitch} et~al.}{2003}]{markevitch2003}
{Markevitch} M.,  {Bautz} M.~W.,  {Biller} B.,  {Butt} Y.,  {Edgar} R.,
  {Gaetz} T.,  {Garmire} G.,  {Grant} C.~E.,  {Green} P.,  {Juda} M.,
  {Plucinsky} P.~P.,  {Schwartz} D.,  {Smith} R.,  {Vikhlinin} A.,  {Virani}
  S.,  {Wargelin} B.~J.,    {Wolk} S.,  2003, \apj, 583, 70

\bibitem[\protect\citeauthoryear{{Meneghetti}, {Argazzi}, {Pace}, {Moscardini},
  {Dolag}, {Bartelmann}, {Li} \& {Oguri}}{{Meneghetti}
  et~al.}{2007}]{meneghetti2007b}
{Meneghetti} M.,  {Argazzi} R.,  {Pace} F.,  {Moscardini} L.,  {Dolag} K.,
  {Bartelmann} M.,  {Li} G.,    {Oguri} M.,  2007, \aap, 461, 25

\bibitem[\protect\citeauthoryear{{Meneghetti}, {Bartelmann} \&
  {Moscardini}}{{Meneghetti} et~al.}{2003a}]{meneghetti2003b}
{Meneghetti} M.,  {Bartelmann} M.,    {Moscardini} L.,  2003a, \mnras, 346, 67

\bibitem[\protect\citeauthoryear{{Meneghetti}, {Bartelmann} \&
  {Moscardini}}{{Meneghetti} et~al.}{2003b}]{meneghetti2003a}
{Meneghetti} M.,  {Bartelmann} M.,    {Moscardini} L.,  2003b, \mnras, 340, 105

\bibitem[\protect\citeauthoryear{{Merten}, {Cacciato}, {Meneghetti}, {Mignone}
  \& {Bartelmann}}{{Merten} et~al.}{2008}]{merten2008}
{Merten} J.,  {Cacciato} M.,  {Meneghetti} M.,  {Mignone} C.,    {Bartelmann}
  M.,  2008, ArXiv e-prints, astro-ph/0806196, 806

\bibitem[\protect\citeauthoryear{{Mewe}, {Gronenschild} \& {van den
  Oord}}{{Mewe} et~al.}{1985}]{mewe1985}
{Mewe} R.,  {Gronenschild} E.~H.~B.~M.,    {van den Oord} G.~H.~J.,  1985,
  \aaps, 62, 197

\bibitem[\protect\citeauthoryear{{Minor} \& {Kaplinghat}}{{Minor} \&
  {Kaplinghat}}{2007}]{minor2007}
{Minor} Q.~E.,  {Kaplinghat} M.,  2007, ArXiv e-prints, astro-ph/07112537, 711

\bibitem[\protect\citeauthoryear{{Miralda-Escude}}{{Miralda-Escude}}{1995}]{mi%
ralda1995}
{Miralda-Escude} J.,  1995, \apj, 438, 514

\bibitem[\protect\citeauthoryear{{Moore}, {Governato}, {Quinn}, {Stadel} \&
  {Lake}}{{Moore} et~al.}{1998}]{moore1998}
{Moore} B.,  {Governato} F.,  {Quinn} T.,  {Stadel} J.,    {Lake} G.,  1998,
  \apjl, 499, L5+

\bibitem[\protect\citeauthoryear{{Morandi} \& {Ettori}}{{Morandi} \&
  {Ettori}}{2007}]{morandi2007b}
{Morandi} A.,  {Ettori} S.,  2007, \mnras, pp 708--+

\bibitem[\protect\citeauthoryear{{Morandi}, {Ettori} \& {Moscardini}}{{Morandi}
  et~al.}{2007}]{morandi2007a}
{Morandi} A.,  {Ettori} S.,    {Moscardini} L.,  2007, \mnras, 379, 518

\bibitem[\protect\citeauthoryear{{Nagai}, {Kravtsov} \& {Vikhlinin}}{{Nagai}
  et~al.}{2007}]{nagai2007}
{Nagai} D.,  {Kravtsov} A.~V.,    {Vikhlinin} A.,  2007, \apj, 668, 1

\bibitem[\protect\citeauthoryear{{Natarajan} \& {Kneib}}{{Natarajan} \&
  {Kneib}}{1997}]{natarajan1997}
{Natarajan} P.,  {Kneib} J.-P.,  1997, \mnras, 287, 833

\bibitem[\protect\citeauthoryear{{Navarro}, {Frenk} \& {White}}{{Navarro}
  et~al.}{1995}]{navarro1995}
{Navarro} J.~F.,  {Frenk} C.~S.,    {White} S.~D.~M.,  1995, \mnras, 275, 720

\bibitem[\protect\citeauthoryear{{Navarro}, {Frenk} \& {White}}{{Navarro}
  et~al.}{1997}]{navarro1997}
{Navarro} J.~F.,  {Frenk} C.~S.,    {White} S.~D.~M.,  1997, \apj, 490, 493

\bibitem[\protect\citeauthoryear{{Ota}, {Pointecouteau}, {Hattori} \&
  {Mitsuda}}{{Ota} et~al.}{2004}]{ota2004}
{Ota} N.,  {Pointecouteau} E.,  {Hattori} M.,    {Mitsuda} K.,  2004, \apj,
  601, 120

\bibitem[\protect\citeauthoryear{{Quadri}, {M{\"o}ller} \&
  {Natarajan}}{{Quadri} et~al.}{2003}]{quadri2003}
{Quadri} R.,  {M{\"o}ller} O.,    {Natarajan} P.,  2003, \apj, 597, 659

\bibitem[\protect\citeauthoryear{{Rasia}, {Ettori}, {Moscardini}, {Mazzotta},
  {Borgani}, {Dolag}, {Tormen}, {Cheng} \& {Diaferio}}{{Rasia}
  et~al.}{2006}]{rasia2006}
{Rasia} E.,  {Ettori} S.,  {Moscardini} L.,  {Mazzotta} P.,  {Borgani} S.,
  {Dolag} K.,  {Tormen} G.,  {Cheng} L.~M.,    {Diaferio} A.,  2006, \mnras,
  369, 2013

\bibitem[\protect\citeauthoryear{{Rasia}, {Tormen} \& {Moscardini}}{{Rasia}
  et~al.}{2004}]{rasia2004}
{Rasia} E.,  {Tormen} G.,    {Moscardini} L.,  2004, \mnras, 351, 237

\bibitem[\protect\citeauthoryear{{Rood}, {Page}, {Kintner} \& {King}}{{Rood}
  et~al.}{1972}]{rood1972}
{Rood} H.~J.,  {Page} T.~L.,  {Kintner} E.~C.,    {King} I.~R.,  1972, \apj,
  175, 627

\bibitem[\protect\citeauthoryear{{Sand}, {Treu} \& {Ellis}}{{Sand}
  et~al.}{2002}]{sand2002}
{Sand} D.~J.,  {Treu} T.,    {Ellis} R.~S.,  2002, \apjl, 574, L129

\bibitem[\protect\citeauthoryear{{Sand}, {Treu}, {Ellis}, {Smith} \&
  {Kneib}}{{Sand} et~al.}{2008}]{sand2008}
{Sand} D.~J.,  {Treu} T.,  {Ellis} R.~S.,  {Smith} G.~P.,    {Kneib} J.-P.,
  2008, \apj, 674, 711

\bibitem[\protect\citeauthoryear{{Sand}, {Treu}, {Smith} \& {Ellis}}{{Sand}
  et~al.}{2004}]{sand2004}
{Sand} D.~J.,  {Treu} T.,  {Smith} G.~P.,    {Ellis} R.~S.,  2004, \apj, 604,
  88

\bibitem[\protect\citeauthoryear{{Schmidt} \& {Allen}}{{Schmidt} \&
  {Allen}}{2007}]{schmidt2007}
{Schmidt} R.~W.,  {Allen} S.~W.,  2007, \mnras, 379, 209

\bibitem[\protect\citeauthoryear{{Shu}, {Zhou}, {Bartelmann}, {Comerford},
  {Huang} \& {Mellier}}{{Shu} et~al.}{2008}]{shu2008}
{Shu} C.,  {Zhou} B.,  {Bartelmann} M.,  {Comerford} J.~M.,  {Huang} J.~.,
  {Mellier} Y.,  2008, ArXiv e-prints, astro-ph/08051148, 805

\bibitem[\protect\citeauthoryear{{Smail}, {Ellis}, {Dressler}, {Couch},
  {Oemler}, {Sharples} \& {Butcher}}{{Smail} et~al.}{1997}]{smail1997}
{Smail} I.,  {Ellis} R.~S.,  {Dressler} A.,  {Couch} W.~J.,  {Oemler} A.~J.,
  {Sharples} R.~M.,    {Butcher} H.,  1997, \apj, 479, 70

\bibitem[\protect\citeauthoryear{{Soucail}, {Covone} \& {Kneib}}{{Soucail}
  et~al.}{2007}]{soucail2007}
{Soucail} G.,  {Covone} G.,    {Kneib} J.-P.,  2007, in {Kissler-Patig} M.,
  {Walsh} J.~R.,   {Roth} M.~M.,  eds, Science Perspectives for 3D Spectroscopy
  {A VIMOS-IFU Survey of z ${\sim}$ 0.2 Massive Lensing Galaxy Clusters:
  Constraining Cosmography}.
pp 181--+

\bibitem[\protect\citeauthoryear{{Stocke}, {Morris}, {Gioia}, {Maccacaro},
  {Schild}, {Wolter}, {Fleming} \& {Henry}}{{Stocke} et~al.}{1991}]{stocke1991}
{Stocke} J.~T.,  {Morris} S.~L.,  {Gioia} I.~M.,  {Maccacaro} T.,  {Schild} R.,
   {Wolter} A.,  {Fleming} T.~A.,    {Henry} J.~P.,  1991, \apjs, 76, 813

\bibitem[\protect\citeauthoryear{{Tozzi}}{{Tozzi}}{2007}]{tozzi2007}
{Tozzi} P.,  2007, in {Papantonopoulos} L.,  ed., The Invisible Universe: Dark
  Matter and Dark Energy Vol.~720 of Lecture Notes in Physics, Berlin Springer
  Verlag, {Cosmological Parameters from Galaxy Clusters: An Introduction}.
pp 125--+

\bibitem[\protect\citeauthoryear{{Vikhlinin}, {Kravtsov}, {Forman}, {Jones},
  {Markevitch}, {Murray} \& {Van Speybroeck}}{{Vikhlinin}
  et~al.}{2006}]{vikhlinin2006}
{Vikhlinin} A.,  {Kravtsov} A.,  {Forman} W.,  {Jones} C.,  {Markevitch} M.,
  {Murray} S.~S.,    {Van Speybroeck} L.,  2006, \apj, 640, 691

\bibitem[\protect\citeauthoryear{{Voigt} \& {Fabian}}{{Voigt} \&
  {Fabian}}{2006}]{voigt2006}
{Voigt} L.~M.,  {Fabian} A.~C.,  2006, \mnras, 368, 518

\bibitem[\protect\citeauthoryear{{Voit}}{{Voit}}{2005}]{voit2005}
{Voit} G.~M.,  2005, Reviews of Modern Physics, 77, 207

\bibitem[\protect\citeauthoryear{{White}, {Jones} \& {Forman}}{{White}
  et~al.}{1997}]{white1997}
{White} D.~A.,  {Jones} C.,    {Forman} W.,  1997, \mnras, 292, 419

\bibitem[\protect\citeauthoryear{{Wilms}, {Allen} \& {McCray}}{{Wilms}
  et~al.}{2000}]{wilms2000}
{Wilms} J.,  {Allen} A.,    {McCray} R.,  2000, \apj, 542, 914

\bibitem[\protect\citeauthoryear{{Wu}}{{Wu}}{1993}]{wu1993}
{Wu} X.-P.,  1993, \apj, 411, 513

\bibitem[\protect\citeauthoryear{{Wu} \& {Fang}}{{Wu} \& {Fang}}{1996}]{wu1996}
{Wu} X.-P.,  {Fang} L.-Z.,  1996, \apjl, 467, L45+

\end{thebibliography}
